\documentclass[reprint, aps,twocolumn,showpacs,showkeys, 10pt, longbibliography, nolinenumbers, floatfix]{revtex4-2}

\usepackage{amsmath}
\usepackage{graphicx}
\usepackage[caption=false]{subfig}
\usepackage{float}
\usepackage{dcolumn}
\usepackage{multirow}
\usepackage{natbib}
\usepackage{microtype}
\usepackage[colorlinks = true,linkcolor = blue,urlcolor  = blue,citecolor = blue,anchorcolor = blue]{hyperref}
\usepackage{enumerate}
\usepackage{bm}
\usepackage{orcidlink}

\begin{document}

\title{Colour-flavour locked quark stars in light of the compact object in the HESS J1731-347 and the GW190814 event}

\author{P.T. Oikonomou\orcidlink{0009-0006-3585-1309}}
\email{pavoikon@auth.gr}

\author{Ch.C. Moustakidis\orcidlink{0000-0003-3380-5131}}
\email{moustaki@auth.gr}

\affiliation{Department of Theoretical Physics, Aristotle University of Thessaloniki, 54124 Thessaloniki, Greece}

\date{7 September 2023}

\begin{abstract}

The central compact object within HESS J1731-
347 possesses unique mass and radius properties
that renders it a compelling candidate for a self-bound
star. In this research, we examine the capability of quark stars composed of colour superconducting quark matter to explain the latter object by using its marginalised posterior distribution and imposing it as a constraint on the relevant parameter space. Namely, we investigate quark matter for $N_f=2,3$ in the colour superconducting phase, incorporating perturbative QCD corrections, and we derive their properties accordingly. The utilised thermodynamic potential of this work possesses an MIT bag model formalism with the parameters being established as flavour-independent. In this instance, we conclude the favour of the 3-flavour over the 2-flavour colour superconducting quark matter and we isolate our interest on the former, without neglecting the possible favour of the latter in a different framework. The parameter space is further confined due to the additional requirement for a high maximum mass ($M_{\text{TOV}} \geq 2.6 M_{\odot}$), accounting for GW$190814$'s secondary companion. Our study places significant emphasis on the speed of sound and the trace anomaly which was proposed as a measure of conformality [Y. Fujimoto \textit{et al.}, \href{https://journals.aps.org/prl/abstract/10.1103/PhysRevLett.129.252702}{Phys. Rev. Lett. 129, 252702 (2022)}]. We conclude that it is possible for colour-flavour locked quark stars to reach high masses without violating the conformal bound or the $\langle \Theta \rangle _{\mu_B} \geq 0$, provided that the quartic coefficient $\alpha_4$ does not exceed an upper limit which depends on the established $M_{\text{TOV}}$. For $M_{\text{TOV}}=2.6 M_{\odot}$, we find that the limit reads $\alpha_4 \leq 0.594$. Lastly, a further study takes place on the agreement of colour-flavour locked quark stars with additional astrophysical objects including the GW$170817$ and GW$190425$ events, followed by a relevant discussion.

\end{abstract}

\maketitle

\section{Introduction}\label{sec1} 

A substantial amount of activity has been focused on Witten's proposal regarding the true ground state of matter \textcolor{blue}{\cite{Witten-1984}} (following Bodmer's important precursor \textcolor{blue}{\cite{Bodmer-1971}}). The hypothesis of quark matter consisting of $u$, $d$ and $s$ quarks, also called strange matter, having an energy per baryon lower than that of nuclear matter and $u$, $d$ quark matter, along with the expectation of de-confined quark matter at high densities \textcolor{blue}{\cite{Ivanenko-1965,Itoh-1970,Collins-1975,Weber-2005,Bielich-2020}}, has lead to the study of an exotic outcome called Strange Stars \textcolor{blue}{\cite{Bielich-2020,Glendenning-2000,Alock-1986,Hanesel-1986}}. Such objects, due to their nature, can reach arbitrarily small radii and masses \textcolor{blue}{\cite{Bielich-2020}}.\

Later work has shown that the aforementioned matter should be a colour superconductor, a degenerate Fermi gas in which the quarks near the Fermi surface form Cooper pairs, breaking the colour gauge symmetry \textcolor{blue}{\cite{Alford-2001a,Alford-2008,Rajagopal-2000}}. At asymptotically high densities, the most favourable state for strange quark matter is the colour-flavour-locked (CFL) phase, a superfluid which furthermore breaks chiral symmetry, where quarks of all three flavours and colours pair with each other in a correlated way \textcolor{blue}{\cite{Alford-2001a,Alford-2008,Rajagopal-2000,Alford-1999}}.\

Color superconductivity can moreover occur in two-flavour quark matter, where $u$ quarks pair with $d$ quarks, resulting in the 2-flavour color superconducting (2SC) phase, which can be energetically favourable regarding nuclear matter under the application of an appropriate stability window. A 2SC phase with the inclusion of unpaired strange quarks (2SC+s) is also possible, though Alford and Rajagopal have concluded the absence of such phases in the interior of compact stars \textcolor{blue}{\cite{Alford-2002}}.\

A recent analysis on the supernova remnant HESS J$1731-347$  suggests that the mass and radius of the central compact object within it are $M=0.77^{+0.20}_{-0.17}M_{\odot}$ and $R=10.4^{+0.86}_{-0.78}$ km respectively \textcolor{blue}{\cite{Doroshenko-2022}}. This estimation is rather intriguing. Considering that a previous analysis by Suwa et al. \cite{Suwa-2018} indicated that the minimum possible mass of a neutron star is $1.17 M_{\odot}$, it raises the question of whether the compact object in HESS J$1731-347$ could be an exotic one rather than a neutron star. Several studies have already examined the possibility of the latter object to be a hybrid star \textcolor{blue}{\cite{Tsaloukidis-2023,Brodie-2023}} or a strange star \textcolor{blue}{\cite{JEHorvath-2023, FDiClemente-2023}}.\

We furthermore turn our attention to the notable GW$190814$ event \textcolor{blue}{{\cite{Abbott-2020a}}} which represents the observation of a merger between a $22.2–24.3 \ M_{\odot}$ black hole and a compact
object with a mass of $2.50–2.67 \ M_{\odot}$ (measurements reported at the $90\%$ credible level). Specifically, the primary component of the GW190814 is conclusively a black hole (BH) with mass $m_1 =23.2^{+1.1}_{-1.0}  \ M_{\odot}$ ,while the nature of the secondary companion $m_2 =2.59^{+0.08}_{-0.09}  \ M_{\odot}$ still remains uncertain due to the absence of measurable tidal deformations and the lack of an electromagnetic counterpart. In this study, we also explore the possibility that the aforementioned object could be a CFL quark star and we investigate the implications of this conjecture on CFL quark matter and the associated parameter space. It is worth emphasizing that the potential for the object to be rapidly rotating has not been ruled out. In any case, the observation of this event can lead both to the improvement of the existing theoretical nuclear models and to the search for new theories that will predict compact systems with such a significant mass.\

In this paper, we use an approximate yet quite accurate thermodynamic potential within an MIT bag model \textcolor{blue}{\cite{Chodos-1974}} formalism, we establish flavour-and-density-independent parameters, to derive an EoS corresponding to CFL quark matter. We follow the respective technique for 2SC quark matter and further produce the stability window on the two-dimensional parameter space $\Delta$ -$B_\text{eff}^{1/4}$ for the aforementioned phases. A conclusion about the energetic preference of CFL over 2SC matter using the specific model is displayed. We narrow our focus on CFL quark matter and proceed to the examination of the speed of sound ($\upsilon_s$) corresponding to the latter, additionally studying the behaviour of the normalised trace anomaly as presented in \textcolor{blue}{\cite{Fujimoto-2022}}. Several studies of quark stars composed of colour-flavour locked quark matter (CFL quark stars) have presented that these objects can reach great maximum masses \textcolor{blue}{\cite{GLugones-2003, JEHorvath-2004, AKurkela-2010, CVFlores-2017}}. Our prime goal is to present the capability of CFL quark stars to explain the HESS J$1731-347$ compact object, as well as objects with masses equal or greater than that of GW$190814$'s secondary companion \textcolor{blue}{\cite{Abbott-2020a}} without violating the conformal limit ($\upsilon_s^2 \rightarrow 1/3$) or the positive trace anomaly bound proposed in \textcolor{blue}{\cite{Fujimoto-2022}}. A depiction on how these two events can constrain the parameter space takes place. Lastly, we investigate the further compliance of CFL quark stars, whose parameters are extracted from the aforementioned confined area, with various astrophysical objects \textcolor{blue}{\cite{Romani-2022,Riley-2021,Antoniadis-2013,Arzoumanian-2018,Nattila-2017,Miller-2019}} and the LIGO-Virgo events \textcolor{blue}{\cite{Abbott-2017,Abbott-2020b}}. We note the utilisation of natural units throughout this research, excluding the radii of the objects in examination which are described in SI units.\

The presented work is organized as follows. Section \ref{sec2} is devoted to the discussion of the properties of colour superconducting quark matter.
In Section~\ref{sec3} we present the confinements on the $\Delta - B_{\text{eff}}$ parameter space induced by observational constraints while Section~\ref{sec4} is dedicated to the presentation of the basic formalism concerning the tidal deformability. In Section~\ref{sec5} we study the behaviour of the trace anomaly within the framework of CFL quark matter and examine the compatibility of the latter with the limits induced by the speed of sound. In Section~\ref{sec6} we present our results, followed by a discussion regarding their implications. Finally, we close with some concluding remarks in Section~\ref{sec7}.

\section{Properties of Colour Superconducting Quark Matter}\label{sec2}
\subsection{CFL Quark Matter}
We begin by displaying the thermodynamic potential of the CFL phase as shown in \textcolor{blue}{\cite{Alford-2003}} 
\begin{eqnarray}
    \Omega_{\text{CFL}}=\Omega^{\text{quarks}}_{\text{CFL}}(\mu) +\Omega^{\text{GB}}_{\text{CFL}}(\mu,\mu_{e})+\Omega^{\text{electrons}}(\mu_e).
    \label{1}
\end{eqnarray}
Where $3\mu=\mu_u+\mu_d+\mu_s$ and $\mu_e$ is the baryon and electron chemical potential, respectively. The presence of the second term in Eq.\textcolor{blue}{~(\ref{1})}, which is the contribution from the Goldstone bosons, arises due to the chiral symmetry breaking induced by the CFL phase\textcolor{blue}{\cite{Alford-1999,Alford-2003,Alford-2005}}.\

Rajagopal \& Wilczek \textcolor{blue}{\cite{Rajagopal-2001}} showed that the CFL phase is electrically neutral with an equal number of $u$, $d$ and $s$ quarks, in spite of the unequal masses, as long as the strange quark mass $m_s$ and the electron chemical potential $\mu_e$ is not too large. In the following, we establish $m_s=95$ MeV \textcolor{blue}{\cite{Workman-2022}} and we neglect the electron contribution in Eq.\textcolor{blue}{~(\ref{1})}. In this case, Alford \& Reddy demonstrated that the second term in Eq.\textcolor{blue}{~(\ref{1})} can be also safely neglected \textcolor{blue}{\cite{Alford-2003}}. Thus, the thermodynamic potential to order $\Delta^2$ including p(erturbative) QCD corrections is \textcolor{blue}{\cite{Alford-2003,Alford-2005,Fraga-2001,Weissenborn-2011,Alford-2001b,Lugones-2002}}

\begin{eqnarray}
    \Omega_{\text{CFL}}&=&\frac{6}{\pi^2} \displaystyle{\int_0^\nu p^2(p-\mu) dp} +\frac{3}{\pi^2} \displaystyle{\int_0^\nu p^2(\sqrt{p^2+m_s^2}-\mu})dp \nonumber \\
    &+& \frac{3\mu^4}{4\pi^2}(1-a_4) - \frac{3\Delta^2\mu^2}{\pi^2} + B_{\text{eff}}.
    \label{2}
\end{eqnarray}

Where $\nu$ is the common Fermi momentum of the quarks that are ``going to pair'', forced by the CFL phase so that charge neutrality is established without the inclusion of an electron number density
\begin{eqnarray}
    \nu=2\mu - \sqrt{\mu^2+\frac{m_s^2}{3}}.
    \label{3}
\end{eqnarray}
Hence, the quark number densities are $n_u=n_d=n_s=(\nu^3 +\mu^3(a_4 - 1) +2\Delta^2\mu)/\pi^2$.\\
\hspace*{5pt} The pQCD corrections are represented by the (1-$a_4$) term up to $\mathcal{O}(a_s^2)$, where the quartic coefficient $a_4$ can vary from the value $a_4=1$ when no strong interactions are taken into consideration, to rather small values when these interactions become important \textcolor{blue}{\cite{Alford-2005,Fraga-2001}}. The next term of Eq.\textcolor{blue}{~(\ref{2})} represents the colour superconductivity contribution via the gap parameter $\Delta$ while the last term is the effective bag constant $B_{\text{eff}}$, representing the non-perturbative QCD vacuum effects.\

By taking into account that $m_s < \mu$, Lugones $\&$ Horvath argued that an approximation of $\Omega_{\text{CFL}}$ up to $m_s^2$ is sufficiently accurate \textcolor{blue}{\cite{Lugones-2002}}. Thus, by expanding the thermodynamic potential in powers of $m_s/\mu$, Eq.\textcolor{blue}{~(\ref{2})} becomes \textcolor{blue}{\cite{Alford-2005,Lugones-2002}}
\begin{eqnarray}
    \Omega_{\text{CFL}}=-\frac{3a_4}{4\pi^2}\mu^4 + \frac{3m^2_s-12\Delta^2}{4\pi^2}\mu^2 +B_{\text{eff}}.
    \label{4}
\end{eqnarray}
By utilising the thermodynamic relations
\begin{align}
    P&=-\Omega, \qquad \qquad n_i=\frac{\partial P}{\partial \mu_i}, \nonumber \\
    {\cal E}&= \displaystyle{\sum_i n_i\mu_i} - P,
    \label{5}
\end{align}
we will derive an EoS that describes the energy density ${\cal E}$ of the CFL state as a function of the respective pressure $P$. For simplicity reasons, we define two new parameters
\begin{align}
    k=\sqrt{a_4}, \qquad w=\frac{a_2}{k},
    \label{6}
\end{align}
where we have introduced the quadratic coefficient $a_2=m^2_s -4\Delta^2$, as displayed by Alford \textit{et al}. \textcolor{blue}{\cite{Alford-2005}}. According to Eqs.\textcolor{blue}{~(\ref{5})}, the energy density reads
\begin{eqnarray}
    {\cal E}_{\text{CFL}} = \frac{9k^2}{4\pi^2}\mu^4 -\frac{3kw}{4\pi^2}\mu^2 +B_{\text{eff}}.
    \label{7}
\end{eqnarray}
 which can be rewritten in terms of the pressure $P$ by using Eq.\textcolor{blue}{~(\ref{4})}
\begin{eqnarray}
    {\cal E}_{\text{CFL}}= 3P +4 B_{\text{eff}} +\frac{3w^2}{4\pi^2} +\frac{w}{\pi} \sqrt{3(P+B_{\text{eff}}) +\frac{9w^2}{16\pi^2}}.
    \label{8}
\end{eqnarray}
Eq.\textcolor{blue}{~(\ref{8})} represents the EoS of strange stars composed of CFL quark matter. An akin result was also derived in \textcolor{blue}{\cite{Zhang-2021}}. We further note that in this instance, the gap condition corresponding to the favouring of CFL quark matter over unpaired neutral strange quark matter was computed to be $\Delta > m^2_s/4\mu$ \textcolor{blue}{\cite{Alford-2002,Alford-2001b}} and was later confined by M. Alford \textit{et al.} to the following stability condition of the CFL phase \textcolor{blue}{\cite{Alford-2004}}
\begin{eqnarray}
    \Delta > \frac{m^2_s}{2\mu}.
    \label{9}
\end{eqnarray}\

By virtue of the Hugenholtz–van Hove theorem \textcolor{blue}{\cite{Bielich-2020}}, the energy per baryon number $E/A$ is equal to the baryon chemical potential $3\mu$ at vanishing pressure (or, equivalently, at $\Omega=0$). Therefore, by using Eq.\textcolor{blue}{~(\ref{4})}, we conclude the following correlation

\begin{eqnarray}
    \left(\frac{E}{A}\right)_{\text{CFL}}={\frac{2\sqrt{6}\pi}{k^{1/2}}}\frac{B_{\text{eff}}^{1/2}}{\sqrt{\sqrt{\frac{16\pi^2}{3}B_{\text{eff}}+w^2}-w}},
    \label{11}
\end{eqnarray}
which is crucial for the stability examination of CFL quark matter.

\subsection{2SC Quark Matter}
The scrutiny of the properties of 2SC quark matter is quite more straightforward regarding the CFL state due to the vanishment of both $u$ and $d$ quark masses. Namely, we establish  $n_d=2n_u$
 to ensure charge neutrality, reaching the following form for the corresponding thermodynamic potential \textcolor{blue}{\cite{Zhang-2021}}
 \begin{eqnarray}
\Omega_{2\text{SC}}=-\frac{[3^{-4/3}(1+2^{4/3})]^{-3} a_4}{4\pi^2}\mu^4-\frac{\Delta^2}{\pi^2}\mu^2 +B_{\text{eff}}.
\label{12}
 \end{eqnarray}
In this instance $3\mu= \mu_u +2\mu_d$. We note here that the condition regarding the favouring of 2SC quark matter over neutral unpaired 2-flavour quark matter is satisfied regardless the value of $\Delta$, since it simply reads $\Delta^2>0$. Thus, the superconducting parameter can have arbitrary (positive) values.\\
\hspace*{5pt} The respective EoS, which can be derived by following the exact same formalism depicted on the above section, reads
\begin{eqnarray}
    {\cal E}_{2\text{SC}}&=&3P+4B_{\text{eff}}  \nonumber \\
    &-&\frac{4\Delta^4}{c\pi^2} \left[\sqrt{\frac{c\pi^2}{\Delta^4}(P+B_{\text{eff}})+1}-1\right],
    \label{13}
\end{eqnarray}
 where $c=[3^{-4/3}(1+2^{4/3})]^{-3}k^2$. The above equation once more corroborates the findings presented in \textcolor{blue}{\cite{Zhang-2021}}. Lastly, the energy per baryon number of 2CS quark matter is equal to
 \begin{eqnarray}
     \left(\frac{E}{A}\right)_{\text{2SC}}=\frac{3\sqrt{2} \pi B^{1/2}_{\text{eff}}}{\sqrt{\sqrt{c\pi^2B_{\text{eff}}+\Delta^4}+\Delta^2}}.
     \label{15}
 \end{eqnarray}
 Eqs.\textcolor{blue}{~(\ref{11})} and\textcolor{blue}{~(\ref{15})} are critical for the establishment of the appropriate stability constraints  and  will be utilised in the next section.

 \section{Stability and Observational Constraints} \label{sec3}
\subsection{Stability Constraints}
\begin{figure}[t]
\includegraphics[width=245pt,height=19pc]{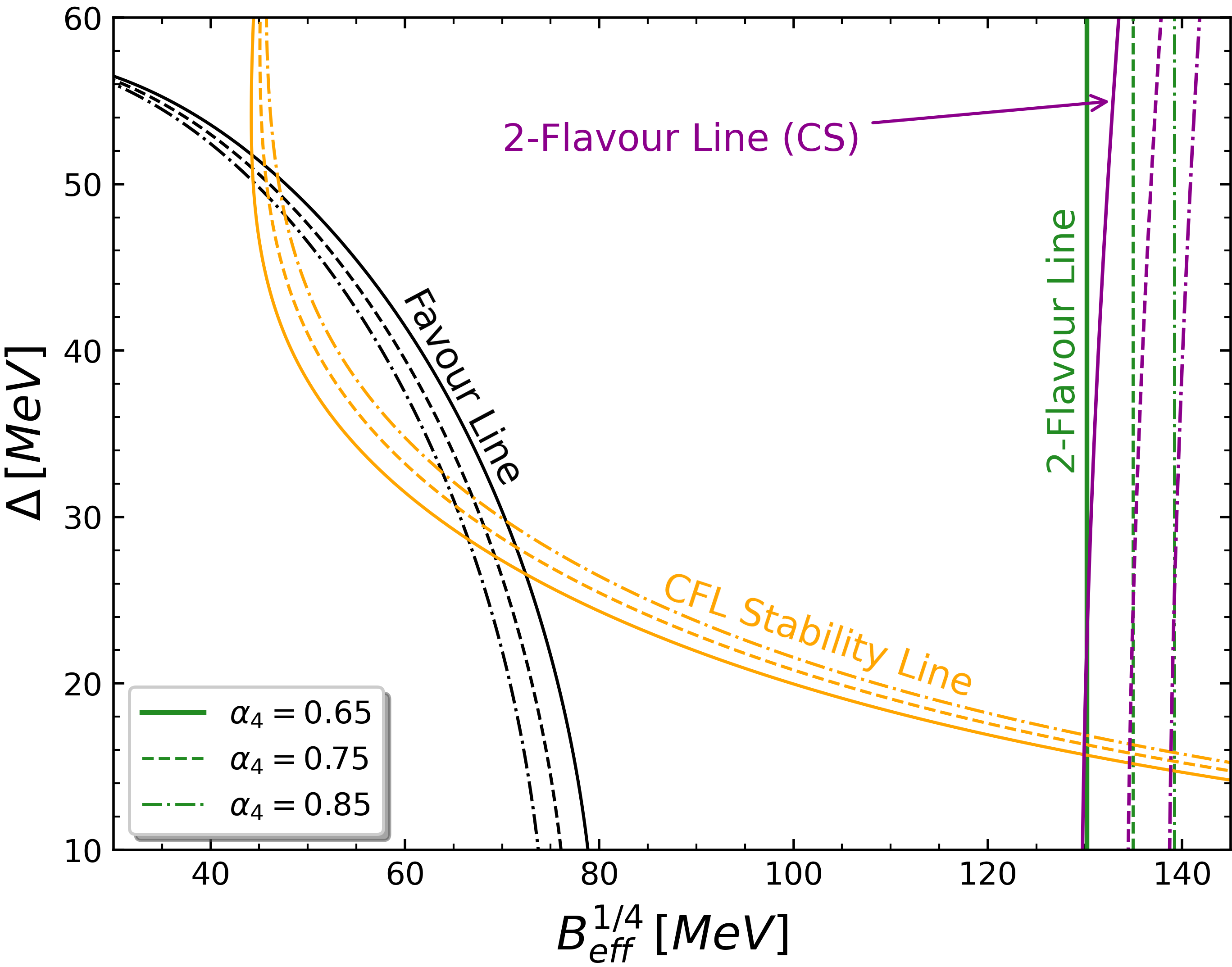}
\caption{The black lines represent the favour boundary, indicating the threshold at which the two investigated states of quark matter are equally stable, illustrated in the $\Delta$-$B_{\text{eff}}^{1/4}$ parameter space. Within the confines of these lines, the 2SC quark matter appear more favourable compared to the CFL quark matter. Conversely, outside this region, the favourability is reversed. The orange lines correspond to the stability condition of the CFL phase, represented by Eq.\textcolor{blue}{~(\ref{9})}. This condition is satisfied within the interior space to the right of these lines. The purple stability lines depict the stability boundaries of 2SC quark matter, with ``CS'' standing for Colour-Superconducting. On the left of these lines, the aforementioned matter is more stable than nuclear matter. Lastly, the green lines serve as a limit for the surety of the favour of ordinary nuclei over neutral unpaired $u,d$ quark matter, making the two-dimensional parameter space left of them forbidden. These results are presented for three distinct $\alpha_4$ values, under the assumption that $m_s = 95 MeV$.\label{fig1}}
\end{figure}

Prior to the study of colour-superconducting quark stars, observational constraints should be taken into account. Particularly, we adopt and establish four stability constraints regarding the quark matter studied above; two of them concern the utter stability of CFL and 2SC quark matter having an energy per baryon $E/A \leq 930$ MeV respectively, where $930$ MeV is roughly the energy per baryon of the most stable nucleus $^{56}\text{Fe}$. The third one concerns the observed stability of ordinary nuclei and the absence of their decay to $u,d$ quark droplets. This term thus corresponds to the energy per baryon of unpaired $u,d$ ($ud$) quark matter and reads 
$(E/A)_{ud}\geq 934$ MeV, where we have added $4$ MeV due to surface effects \textcolor{blue}{\cite{Farhi-1984}}. The latter constrain in conjunction with the second one ($(E/A)_{\text{2SC}} \leq 930$ MeV) induces a tighter lower limit on the gap $\Delta$ range of the 2SC quark matter, depending on the quatric coefficient $a_4$. Particularly, for a specific gap value and lower, the respective range of $B_{\text{eff}}$ values prompted by the second stability constraint violates the stability of the nuclei. As shown in Table\textcolor{blue}{~\ref{tab1}} and Fig.\textcolor{blue}{~\ref{fig1}} (purple lines), we interestingly deduce that as $a_4$ rises, that is as strong interactions become more frail, the lower limit of the gap rises as well. We note that in this work we account for a wide space of the quartic coefficient $\alpha_4 \in [0.45,1]$ \textcolor{blue}{\cite{Alford-2005,Fraga-2001,Zhou-2018,Miao-2021}}.
\begin{table}[b]
\caption{Minimum possible values of the superconducting gap $\Delta$ corresponding to 2SC quark matter, induced by the 2-flavour stability line, for various values of the quartic coefficient $a_4$ and for fixed $m_s=95$ MeV.\label{tab1}}%
\tabcolsep=30pt%
\begin{ruledtabular}
\begin{tabular}{cc}
\multicolumn{2}{c}{\textbf{2SC Quark Matter}}\\
$a_4$ & $\Delta$ (MeV) \\ \hline

$0.45$ & $>18.647$ \\
$0.65$ & $>22.411$ \\
$0.85$ & $>25.628$  \\
$1.00$ & $>27.798$  \\
\end{tabular}
\end{ruledtabular}
\end{table}

The last stability constraint is induced by Eq.\textcolor{blue}{~(\ref{9})} which regards the stability of the CFL phase. Its limitations are presented in Fig.\textcolor{blue}{~\ref{fig1}} (orange lines), where it is concluded that for gaps about $\Delta \approx 16$ MeV ($15$ MeV for  $\alpha_4 = 0.45$) and higher, the parameter space is not further constricted. This is due to the fact that the corresponding stability line is active beyond the third constraint which concerns the stability of nuclei and thus cannot be violated (green lines). Moreover, by accounting for the first stability constraint regarding CFL quark matter ($(E/A)_{\text{CFL}} \leq 930$) along with the confinement of Eq.\textcolor{blue}{~(\ref{9})}, we conclude that the superconducting gap must not recede into values smaller than $14.556$ MeV, so that $\mu|_{P=0} \leq 310$ MeV. This limitation can be simply written in a way that highlights the $m_s$ dependency as $\Delta > m^2_s/620 \, \, \text{MeV}^{-1}$. Note that these results were deduced with the establishment of a density-independent superconducting gap $\Delta$.\

\begin{figure}[b]
\includegraphics[width=245pt,height=19pc]{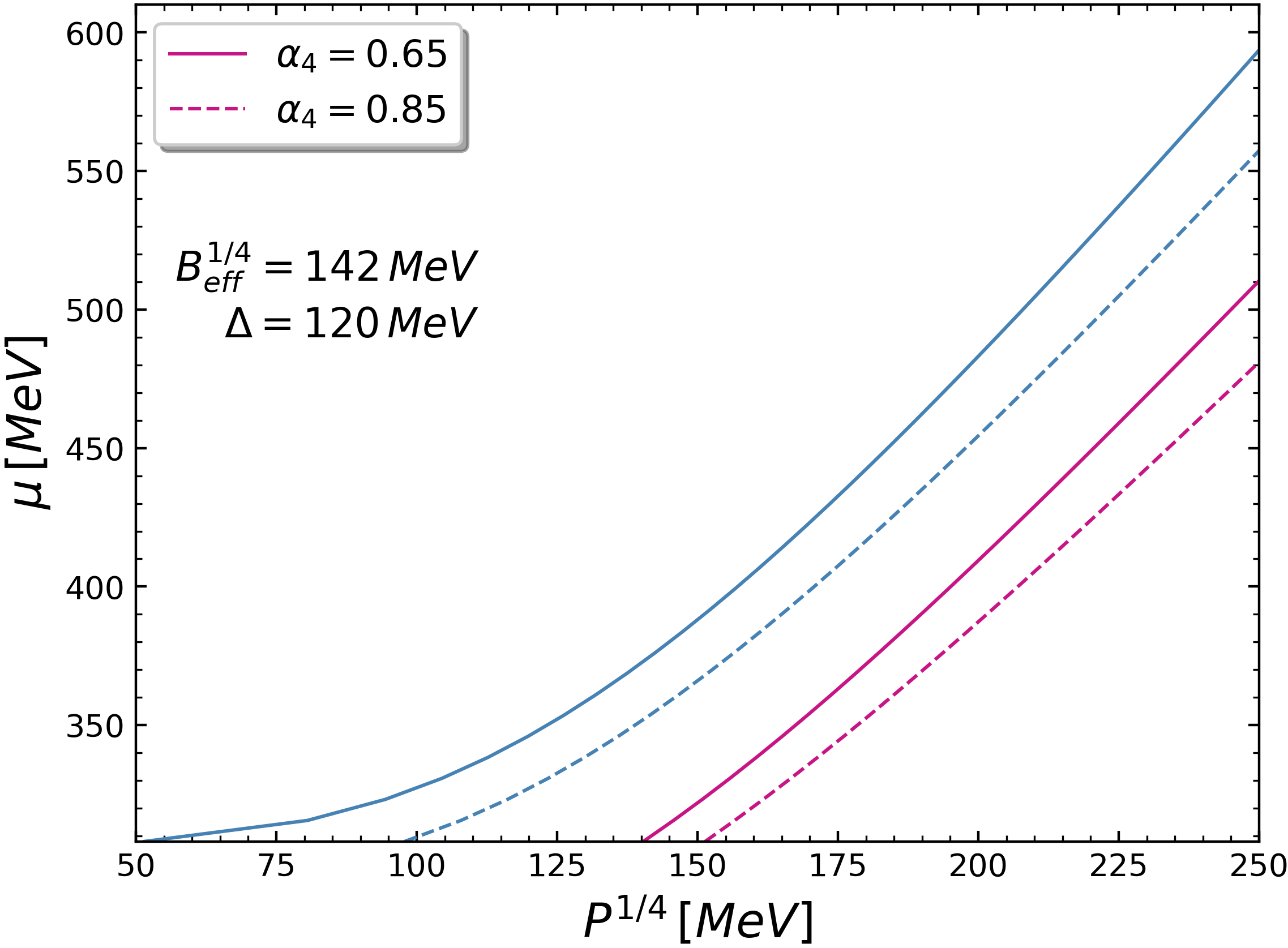}
\caption{The average quark chemical potential as a function of the pressure for the CFL (purple) and 2SC (blue) conditions, at fixed $\Delta$, $B_{\text{eff}}$ and $m_s=95$ MeV and for two different values of $a_4$.\label{fig2}}
\end{figure}

We now direct our study to the examination of the most preferred  quark matter phase and we conclude that the model which is utilised in this work predicts the favouring of CFL over 2SC quark matter. To elaborate, we do find a window in the interior of which 2SC quark matter is more stable than CFL quark matter. However, by taking into account the third constraint which concerns the stability of nuclei, we deduce that the aforementioned area is forbidden since we have to violate the nuclei stability in order to utilise $\Delta-B_{\text{eff}}$ values that allow for the favour of 2SC over CFL matter. This argument is illustrated with clarity in Fig.\textcolor{blue}{~\ref{fig1}} (black lines) and further justified in Fig.\textcolor{blue}{~\ref{fig2}}, where we observe that for a given pressure $P$, the average quark chemical potential $\mu$ with the highest value corresponds to the 2SC state, making it less favourable than the CFL phase. It is evident however that this justification applies on the model in usage where we have established flavour-independent parameters. Flavour-dependence can certainly be possible (e.g. on $B_{\text{eff}}$ \textcolor{blue}{\cite{Buballa-1999,Holdom-2018}}, or $\Delta$ \textcolor{blue}{\cite{Alford-2008,Rajagopal-2000,Alford-1999b,Schafer-2000}}) and affect our depicted results \textcolor{blue}{\cite{QWang-2020}}, thereby indicating new possibilities regarding the true ground state of matter. Without neglecting the possibility of a favoured 2SC quark matter, which can be possible in a more realistic density-and-flavour dependent model, we focus our work solely on CFL quark matter.

\subsection{Observational Constraints}
\begin{figure}[ht]
\includegraphics[width=245pt,height=19pc]{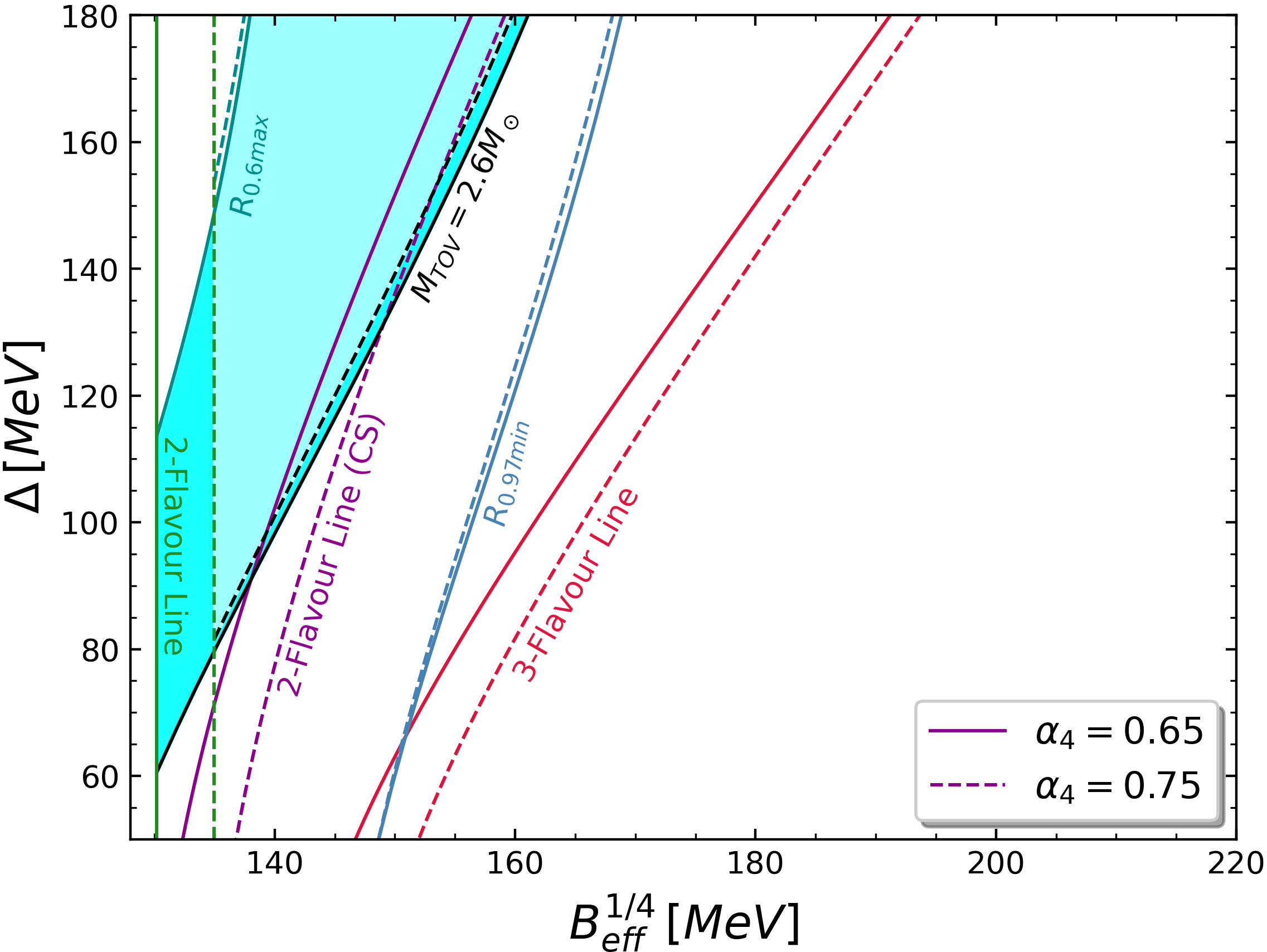}
\caption{The 2-D parameter space $\Delta$-$B_{\text{eff}}^{1/4}$ with applied observational constraints for two different values of $\alpha_4$ and fixed $m_s=95$ MeV. On the left of every stability line, the corresponding matter is more stable than nuclear matter while the astrophysical bounds indicate the appropriate regions that allow for the explanation of the respective object as CFL quark stars; black for the GW$190814$ event ($M_{\text{TOV}}=2.6 M_{\odot}$) and blue for the marginalised constraints of the HESS $J1731$-$347$ remnant ($R_{0.97\text{min}}=9.62$ km and $R_{0.6\text{max}}=11.26$ km). The vivid light blue area represents the window of the parameter space inside of which both of the aforementioned objects can be explained for $\alpha_4=0.65$, while the frail shade corresponds to both ($\alpha_4=0.65$ and $\alpha_4=0.75$) cases. The green lines represent the same limit as the one in Fig.\textcolor{blue}{~\ref{fig1}} and thus the space left of them is excluded.\label{fig3}}
\end{figure}

 In order to investigate the compatibility of CFL quark stars with the HESS J1731-347 remnant and the secondary companion of the GW$190814$ event, additional constraints need to be imposed on the parameter space associated with the measurements of these objects. Specifically, we narrow our focus on the marginalised posterior distribution of HESS J$1731$-$347$'s central compact object (excluding all additional priors described on the text) which reads $M=0.77^{+0.20}_{-0.17}M_{\odot}$ and $R=10.4^{+0.86}_{-0.78}$ km \textcolor{blue}{\cite{Doroshenko-2022}}. By performing a series of cubic spline interpolations we imported robust limitations on the parameter space, corresponding to the latter object, by establishing $R_{0.97\text{min}}=9.62$ km and $R_{0.6\text{max}}=11.26$ km on the radii of CFL quark stars, producing the corresponding lines in Fig.\textcolor{blue}{~\ref{fig3}}. This particular choice of constraints is guided by the form of the $M-R$ curves, depicted in Fig. \ref{fig7} of Sec. \ref{sec6}. It is evident that the lower right ($R_{0.6}=11.26$ km) and upper left ($R_{0.97}=9.62$ km) limits of the aforementioned posterior distribution (indicated by the yellow area in the figure) exclusively constrain the $M-R$ curves of CFL quark stars. By further considering the $M-R$ graph behaviour, which is determined by Eq.\textcolor{blue}{~(\ref{8})}, we have ensured the validity of this formalism with no exceptions. Furthermore, we perform additional interpolations for the constraints related to the GW$190814$ event which we translate to a maximum mass bound of $M_{\text{TOV}} \geq 2.6 M_{\odot}$, producing once more the corresponding lines on the parameter space. Due to the finding of \textcolor{blue}{\cite{Alford-2005,Fraga-2001,Miao-2021}}, we take as an example two different cases of the quartic coefficient $a_4=0.65, 0.75$ and by incorporating the stability constraints of Sec. \textcolor{blue}{\ref{sec3}A}, we depict our overall results in Fig.\textcolor{blue}{~\ref{fig3}} (see comparable figures in \textcolor{blue}{\cite{Weissenborn-2011,Bombaci-2021}}). Note that the gap required in order to describe both of these events is higher than $\Delta=60$ MeV for the $a_4=0.65$ case and higher than $\Delta=80$ MeV for $a_4=0.75$, placing the utility of the respective EoS in the negative $w$ space. The preference of positive over negative $w$ values is a key conclusion of this study and is primarily addressed in Sec. \ref{sec5}.

\section{Tidal Deformability}\label{sec4}

A very important source for the gravitational wave detectors is the gravitational waves from the late phase of the inspiral of a binary compact star system, before the merger \textcolor{blue}{\cite{Postnikov-2010,Flanagan-2008,Hinderer-2008}}. This kind of source leads to the measurement of various properties of compact stars. In the inspiral phase the tidal effects can be detected \textcolor{blue}{\cite{Flanagan-2008}}.

The $k_2$ parameter, also known as tidal Love number, depends on the equation of state and describes the response of a compact star to the tidal field $E_{ij}$~\cite{Flanagan-2008}. The exact relation is given below
\begin{equation}
Q_{ij}=-\frac{2}{3}k_2\frac{R^5}{G}E_{ij}\equiv- \lambda E_{ij},
\label{Love-1}
\end{equation}
where $R$ and $\lambda=2R^5k_2/3G$ is the star's radius and tidal deformability respectively \textcolor{blue}{\cite{Binnington-2009,Damour-1992,Damour-2009}}. The latter, for general purposes, can be rewritten in a dimensionless form as follows
\begin{equation}
    \Lambda=\frac{2}{3}k_2 \left(\frac{R}{GM}\right)^5=\frac{2}{3}k_2 \beta ^{-5},
\end{equation}
with $\beta=GM/R$ being the compactness of the star. The tidal Love number $k_2$ is given by \textcolor{blue}{\cite{Flanagan-2008,Hinderer-2008}}
\begin{eqnarray}
k_2&=&\frac{8\beta^5}{5}\left(1-2\beta\right)^2\left[2-y_R+(y_R-1)2\beta \right]\nonumber\\
& \times&
\left[\frac{}{} 2\beta \left(6  -3y_R+3\beta (5y_R-8)\right) \right. \nonumber \\
&+& 4\beta^3 \left.  \left(13-11y_R+\beta(3y_R-2)+2\beta^2(1+y_R)\right)\frac{}{} \right.\nonumber \\
&+& \left. 3\left(1-2\beta \right)^2\left[2-y_R+2\beta(y_R-1)\right] {\rm ln}\left(1-2\beta\right)\right]^{-1}_,
\label{k2-def}
\end{eqnarray}
where the parameter $y_R$ is evaluated at the star's surface by solving numerically the following differential equation
\begin{align}
r\frac{dy(r)}{dr}+y^2(r)+y(r)F(r)+r^2Q(r)=0, 
\label{D-y-1}
\end{align}
with the initial condition $ y(0)=2$ \textcolor{blue}{\cite{Hinderer-2010}}. $F(r)$ and $Q(r)$ are functions of the energy density ${\cal E}(r)$, pressure $P(r)$, and mass $M(r)$ defined as \textcolor{blue}{\cite{Postnikov-2010}}
\begin{equation}
F(r)=\left[ 1- 4\pi r^2 G \left({\cal E} (r)-P(r) \right)\right]\left(1-\frac{2M(r)G}{r}  \right)^{-1},
\label{Fr-1}
\end{equation}
and
\begin{align}
r^2Q(r)&=4\pi r^2 G \left[5{\cal E} (r)+9P(r)+\frac{{\cal E} (r)+P(r)}{\partial P(r)/\partial{\cal E} (r)}\right]
\nonumber\\
&\times
\left(1-\frac{2M(r)G}{r}  \right)^{-1}- 6\left(1-\frac{2M(r)G}{r}  \right)^{-1} \nonumber \\
&-\frac{4M^2(r)G^2}{r^2}\left(1+\frac{4\pi r^3 P(r)}{M(r)}   \right)^2\left(1-\frac{2M(r)G}{r}\right)^{-2}.
\label{Qr-1}
\end{align}
Eq.\textcolor{blue}{(\ref{D-y-1})} must be solved numerically and self-consistently with the Tolman - Oppenheimer - Volkoff (TOV) equations under the following boundary conditions: $y(0)=2$, $P(0)=P_c$ ($P_{c}$ denotes the central pressure), and $M(0)=0$ \textcolor{blue}{\cite{Postnikov-2010,Hinderer-2008}}. From the numerical solution of the TOV equations, the mass $M$ and radius $R$ of the star can be computed, while the corresponding solution of the differential Eq.\textcolor{blue}{~(\ref{D-y-1})} provides the value of $y_R=y(R)$, where the correction term $-4\pi R^3 {\cal E}_s/M$ \textcolor{blue}{\cite{Bielich-2020,Zhou-2018,Postnikov-2010,Damour-2009,Hinderer-2010,Li-2021,Lourenco-2021}} (${\cal E}_s$ represents the energy density at the surface of the star) must be added which accounts for the energy discontinuity since our focus of interest lies on quark stars. The last parameter along with the quantity $\beta$ are the  basic ingredients  of the tidal Love number $k_2$. Lastly, we present the average tidal deformability of a binary system which is defined as \textcolor{blue}{\cite{Bielich-2020}}
\begin{align}
    \Tilde{\Lambda}=\frac{16}{13} \frac{(M_1 +12 M_2)M_1^4 \Lambda_1 +(M_2 +12 M_1)M_2^4 \Lambda_2}{(M_1 + M_2)^5}.
\end{align}
By exploiting the equations describing the tidal deformability and the Love number $k_2$, we can gain valuable knowledge about the tidal properties
of a star or/and a binary system and compare them with the restrictions induced by several gravitational wave events.

\begin{figure*}[t]

\subfloat{%
\includegraphics[height=7cm,width=1\columnwidth]{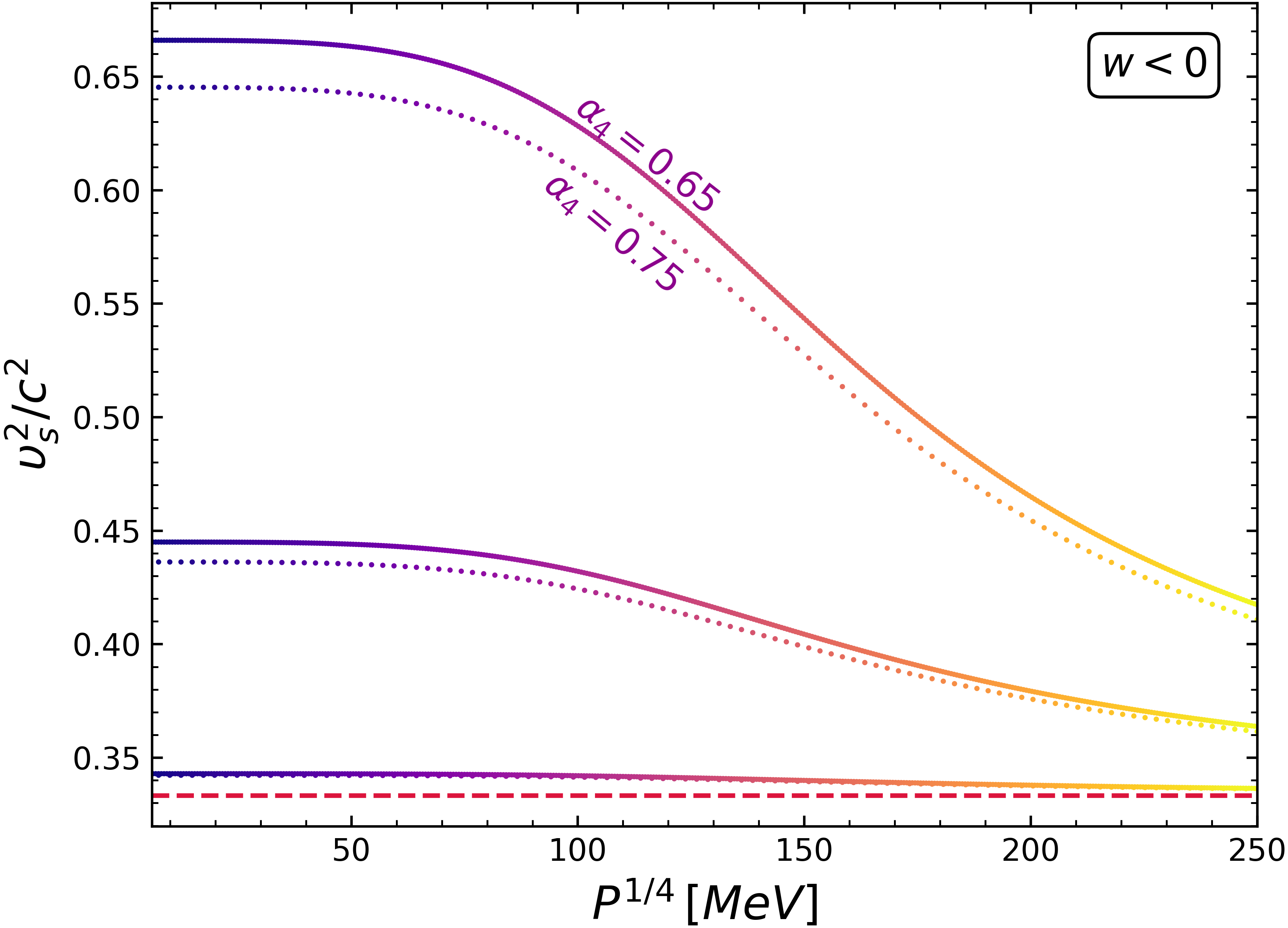}%
}
\subfloat{%
\includegraphics[height=7cm,width=1\columnwidth]{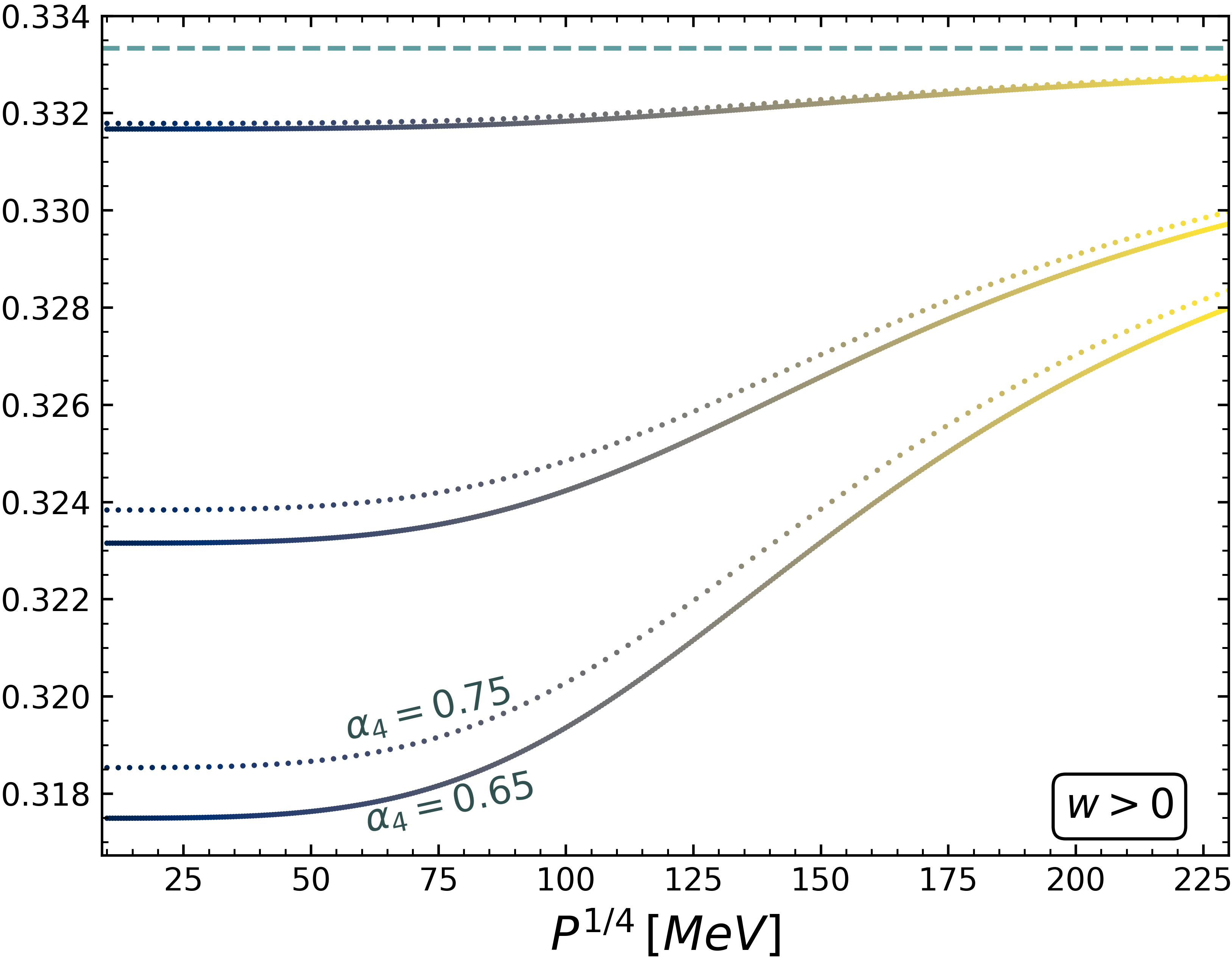}%
}\\
\subfloat{\includegraphics[height=7cm,width=1\columnwidth]{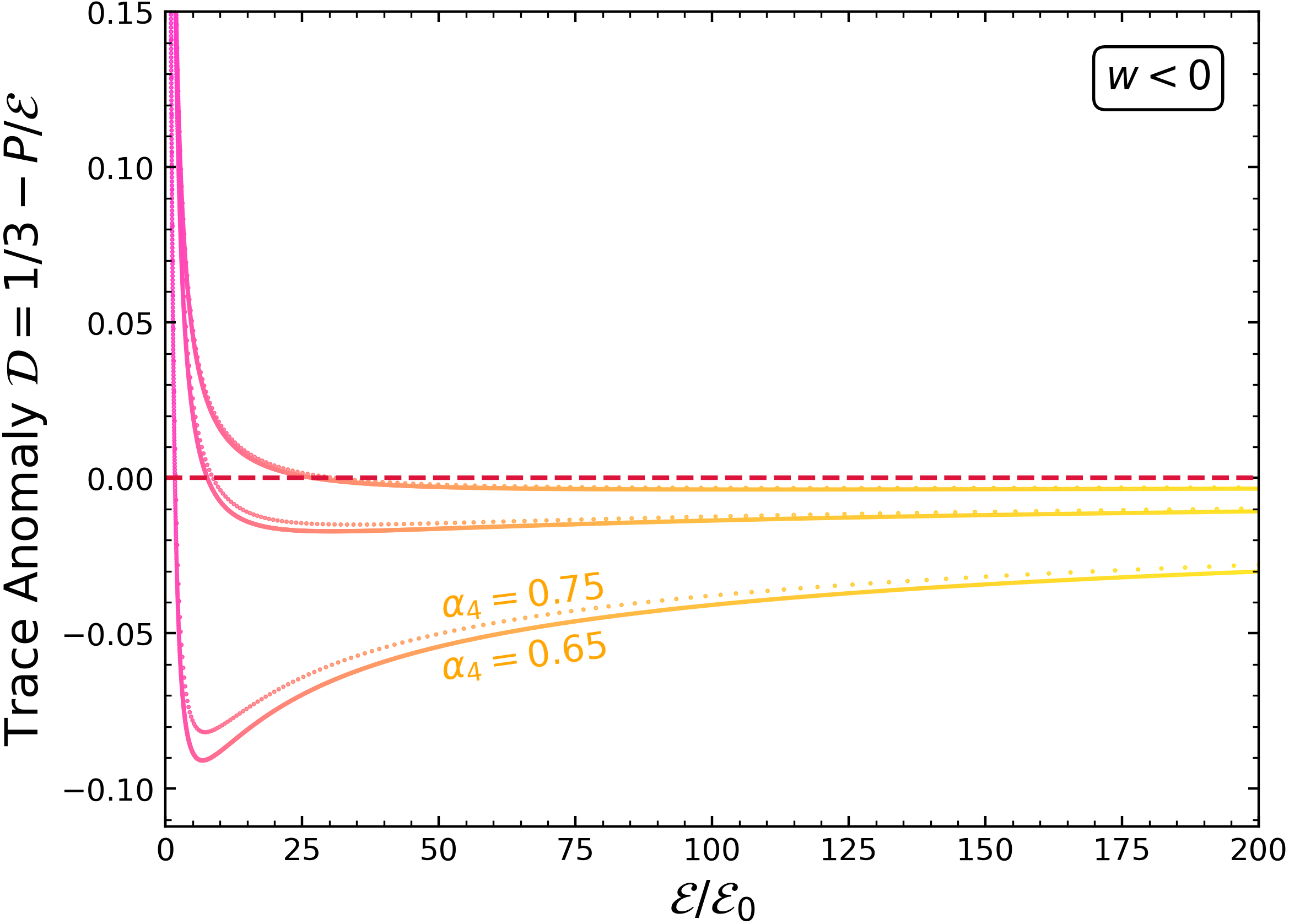}}
\subfloat{\includegraphics[height=7cm,width=1\columnwidth]{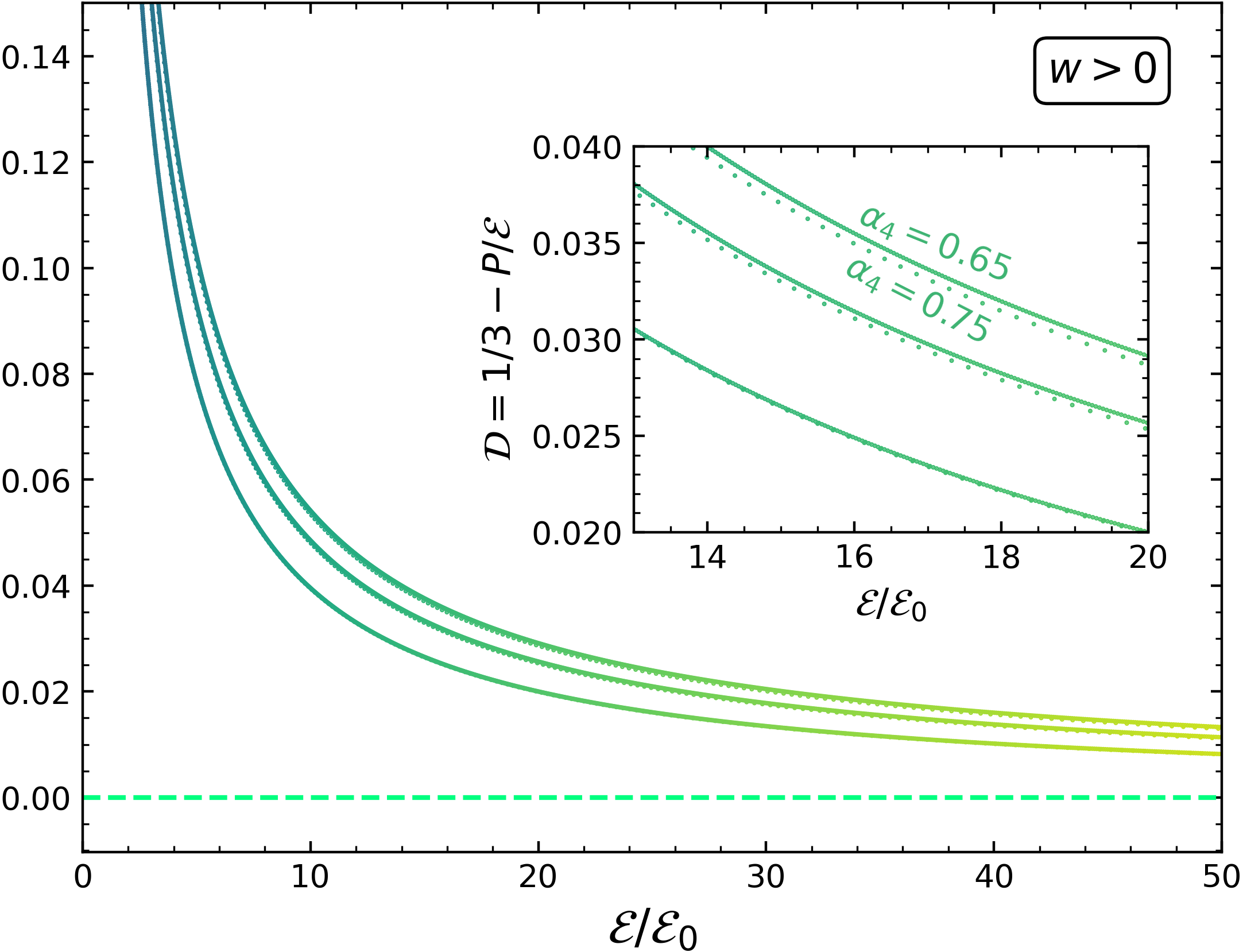}}

\caption{Upper panel: The speed of sound as a function of the pressure in the $w<0$ space (left) for \{$\Delta=60$ MeV, $B_{\text{eff}}^{1/4}=148$ MeV\}, \{$\Delta=120$ MeV, $B_{\text{eff}}^{1/4}=143$ MeV\} and \{$\Delta=180$ MeV, $B_{\text{eff}}^{1/4}=135$ MeV\} along with the $w>0$ case (right) for \{$\Delta=16$ MeV, $B_{\text{eff}}^{1/4}=135$ MeV\}, \{$\Delta=30$ MeV, $B_{\text{eff}}^{1/4}=140$ MeV\} and \{$\Delta=45$ MeV, $B_{\text{eff}}^{1/4}=145$ MeV\}. The pairs in each case are depicted from bottom to top respectively and the dashed lines represent the same limit $\upsilon_s^2 \rightarrow 1/3$. Lower panel: The trace anomaly as a function of the energy density (normalised to the value of the saturation energy density ${\cal E}_0^{1/4}=184.25$ MeV), in the same w areas for \{$\Delta=85$ MeV, $B_{\text{eff}}^{1/4}=135.84$ MeV\}, \{$\Delta=120$ MeV, $B_{\text{eff}}^{1/4}=143$ MeV\} and \{$\Delta=180$ MeV, $B_{\text{eff}}^{1/4}=138$ MeV\} in $w<0$ and \{$\Delta=16$ MeV, $B_{\text{eff}}^{1/4}=140$ MeV\}, \{$\Delta=30$ MeV, $B_{\text{eff}}^{1/4}=138$ MeV\},\{$\Delta=45$ MeV, $B_{\text{eff}}^{1/4}=135$ MeV\} in $w>0$. The dashed lines in this case represent the limit induced from the $\langle \Theta \rangle _{\mu_B} \geq 0$ bound while the pairs are depicted from top to bottom. Each presented pair satisfies the stability constraints of Sec.\textcolor{blue}{\ref{sec3}} while the utilised $w<0$ pairs of the lower panel additionally lie within the frail blue window of Fig.\textcolor{blue}{~\ref{fig3}}. \label{fig4}} 
\end{figure*}

\begin{figure}[t]
\includegraphics[width=245pt,height=19pc]{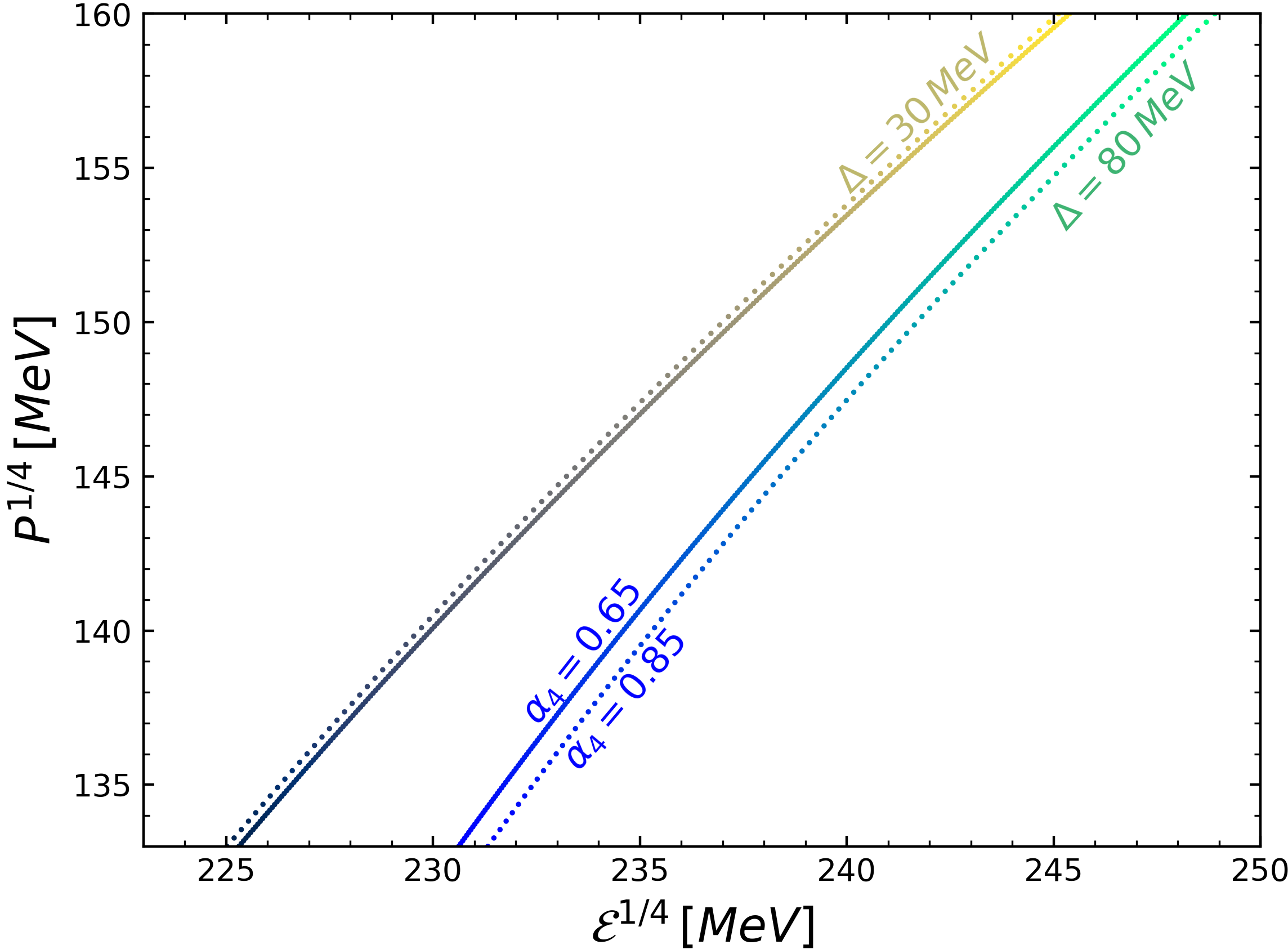}
\caption{The pressure as a function of the energy density for two different values of the superconducting gap $\Delta=30$ MeV ($B_{\text{eff}}^{1/4}=140$ MeV) and $\Delta=80$ MeV ($B_{\text{eff}}^{1/4}=153$ MeV), representing the $w>0$ and $w<0$ space respectively, and for two different values of $\alpha_4$. Note that for $w>0$, an increase of $\alpha_4$ implies a stiffening to the EoS (as the pressure for a given energy density value is higher), while the contrary takes place in the $w<0$ space.\label{fig5}}
\end{figure}

\begin{figure}[t]
\subfloat{
\includegraphics[height=13pc,width=1\columnwidth]{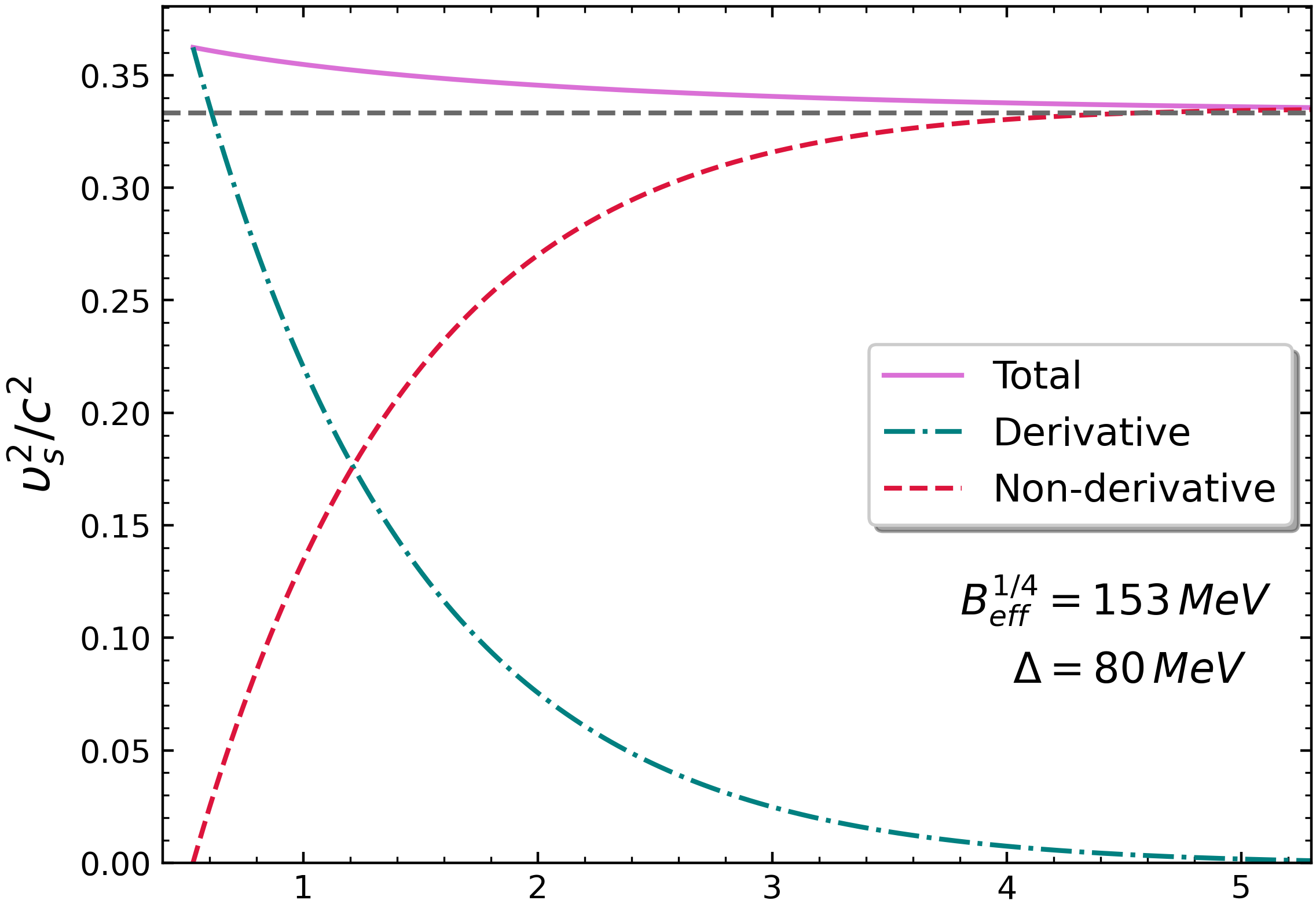}}\\
\subfloat{
\includegraphics[height=13pc,width=1\columnwidth]{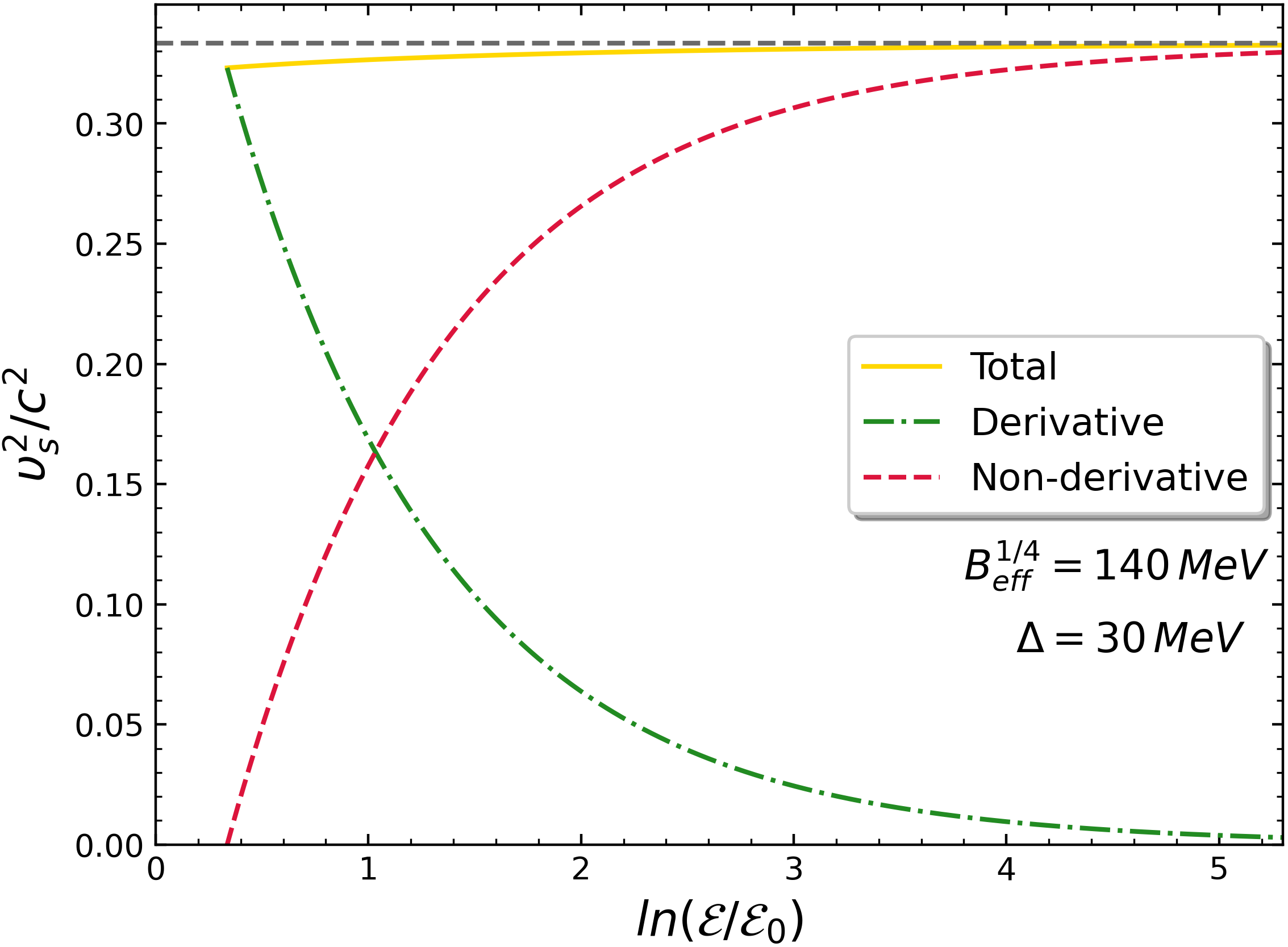}}
\caption{The speed of sound along with the respective derivative and non-derivative terms, as functions of the normalised logarithmic energy density ln$({\cal E}/{\cal E}_0)$ for $w<0$ (upper figure) and $w>0$ (lower figure) with fixed $\alpha_4=0.65$.} \label{fig6}
\end{figure}

\section{Speed of sound and further constraints from the trace anomaly}\label{sec5}
In this section, the scrutiny of the speed of sound is presented in response to the causality ($\upsilon_s \leq 1$) and the conformal ($\upsilon_s \leq 1/\sqrt{3}$) limit. The aforementioned quantity is defined as
\begin{align}
 \frac{1}{\upsilon ^2_s}=\frac{d \cal E}{dP}=3 +  \frac{3w}{2\pi \sqrt{3(P+B_{\text{eff}})+\frac{9w^2}{16\pi^2}}},  \label{23}
\end{align}
and it is evident that in the $\alpha_2 \geq 0$ ($w \geq 0$) space, neither of the limits is violated (however, the  $\upsilon^2_s$ values in this instance are close to the conformal bound as presented in Fig.\textcolor{blue}{~\ref{fig4}}), whereas in the $\alpha_2 < 0$ space ($w < 0$), the speed of sound does violate the conformal limit $\forall$ $B_{\text{eff}}>0$ and $w<0$ (see also the work conducted in \textcolor{blue}{\cite{Bedaque-2015,Tews-2018,Traversi-2022,Moustakidis-2017,Roupas-2021,Ecker-2022,Altiparmak-2022,Hippert-2021,Blaschke-2022,Annala-2023}}). The satisfaction of the causality limit for the latter case can be discerned if one brings Eq.\textcolor{blue}{~(\ref{23})} to the following form
\begin{align}
    P+ B_{\text{eff}}=\frac{3w^2}{4\pi^2}\left(  \frac{1}{(3-\upsilon^{-2}_s)^2} -\frac{1}{4}\right),
\end{align}
and demand that $P+B_{\text{eff}}>0$. Our results seem to be in agreement with the findings of \textcolor{blue}{\cite{Miao-2021}} and are further depicted on Fig.\textcolor{blue}{~\ref{fig4}} which indicates an intriguing behaviour of CFL quark matter; for every case ($w<0$ or $w>0$) we chose to depict three different $\Delta-B_{\text{eff}}^{1/4}$ pairs. For $w<0$, the pairs presented from bottom to top are \{$\Delta=60$ MeV, $B_{\text{eff}}^{1/4}=148$ MeV\}, \{$\Delta=120$ MeV, $B_{\text{eff}}^{1/4}=143$ MeV\} and \{$\Delta=180$ MeV, $B_{\text{eff}}^{1/4}=135$ MeV\} respectively. Whereas for $w>0$, the pairs are \{$\Delta=16$ MeV, $B_{\text{eff}}^{1/4}=135$ MeV\}, \{$\Delta=30$ MeV, $B_{\text{eff}}^{1/4}=140$ MeV\} and \{$\Delta=45$ MeV, $B_{\text{eff}}^{1/4}=145$ MeV\} respectively. Notice now that for the former case ($w<0$), by increasing $\alpha_4$ the speed of sound depletes, whereas for the latter case ($w>0$), the rise of $\alpha_4$ implies, by contrast, an increase to the speed of sound. This behaviour is also present in the stiffness of the respective EoS (as illustrated in Fig.\textcolor{blue}{~\ref{fig5}}), where the first case corresponds to the depletion of it while the second to an increase. One can verify these results by bringing Eq.\textcolor{blue}{~(\ref{8})} to a form $P=P(\cal E)$ with the help of Eqs.\textcolor{blue}{~(\ref{5})} and notice the reduction and increment of the pressure $P$ (for a given ${\cal E}, \Delta $ and $B_{\text{eff}}$) correspondingly. We elucidate that this, however, is not the case for the other two free parameters ($\Delta$ and $B_{\text{eff}}$) as they demonstrate a defined relation with the stiffness of the EoS, independent of the parameter $w$ ($dP/dB_{\text{eff}}<0$ and $dP/d\Delta >0$) \textcolor{blue}{\cite{Lugones-2002,Comment-1}}.\

A recently conducted research \textcolor{blue}{\cite{Fujimoto-2022,Marczenko-2023}}, which proposes the trace anomaly as a measure of conformality, demonstrated that the speed of sound in the case of neutron stars does not violate the conformal bound when $\upsilon_s^2 > 1/3$, but it displays a steep approach to the conformal limit. In particular, the speed of sound $\upsilon_s$ is expressed solely in terms of the trace anomaly and it is suggested that the latter is a more comprehensive quantity than $\upsilon_s$. The normalised trace anomaly is defined as
\begin{equation}
{\cal D} \equiv \frac{\langle \Theta \rangle _{T, \mu_B}}{3\cal E}=\frac{1}{3}-\frac{P}{\cal E}, \qquad \langle \Theta \rangle _{T, \mu_B}={\cal E}-3P,
\label{trace-1}
\end{equation}
where we replaced the original representation letter $\Delta$ with $\cal D$ to avoid any possible confusion with the superconducting gap parameter. In this instance, $\langle \Theta \rangle _{T, \mu_B}$ is the matter part of the trace of the energy-momentum tensor and the quantity ${\cal D}$ must satisfy the obvious constraints $-2/3 \leq {\cal D}  \leq 1/3$. The speed of sound can then be written in terms of ${\cal D}$ as follows
\begin{equation}
\upsilon_s^2=\frac{1}{3}-{\cal D}-{\cal E}\frac{d {\cal D}}{d {\cal E}},
\label{speed-1}
\end{equation}
where $1/3 -{\cal D}$ and $-{\cal E} d{\cal D}/d{\cal E}$ is the non-derivative and derivative term respectively. The latter is responsible for the peak of the speed of sound as shown in \textcolor{blue}{\cite{Fujimoto-2022}}. It is furthermore implied that there may be a positive bound of the trace anomaly reading $\langle \Theta \rangle _{\mu_B} \geq 0$, for cold dense matter. We examine this limit and find that for $w<0$ a violation is possible if the pressure of CFL quark matter infringes the following condition
\begin{align}
    B^2_{\text{eff}} \geq \frac{3w^2}{16\pi^2} (P - B_{\text{eff}}).
\end{align}
However, it is apparent that for $w \geq 0$ the positive bound is not violated. These results are presented on the lower panel of Fig.\textcolor{blue}{~\ref{fig4}} for the trace anomaly $\mathcal{D}$ as a function of the energy density $\cal E$. A further depiction of the derivative and non-derivative parts of the speed of sound is displayed on Fig.\textcolor{blue}{~\ref{fig6}}  where we have included great densities to exhibit the respective behaviour of $\upsilon_s$, asymptotically reaching the $1/3$ value as predicted from Eq.\textcolor{blue}{~(\ref{23})}.

\begin{figure*}[t]
\subfloat[]{
\includegraphics[height=10cm,width=1\columnwidth]{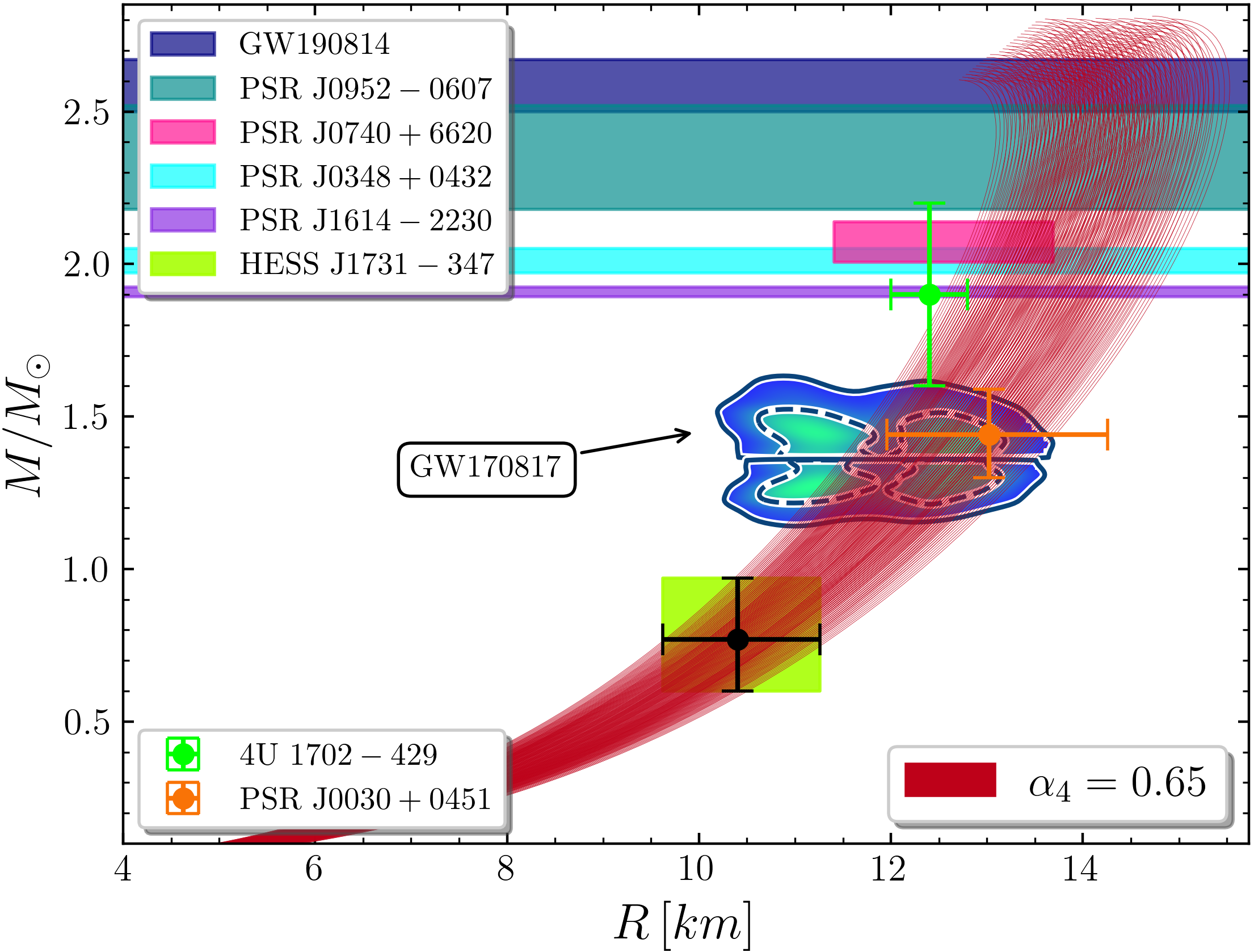}
}
\subfloat[]{
\includegraphics[height=10cm,width=1\columnwidth]{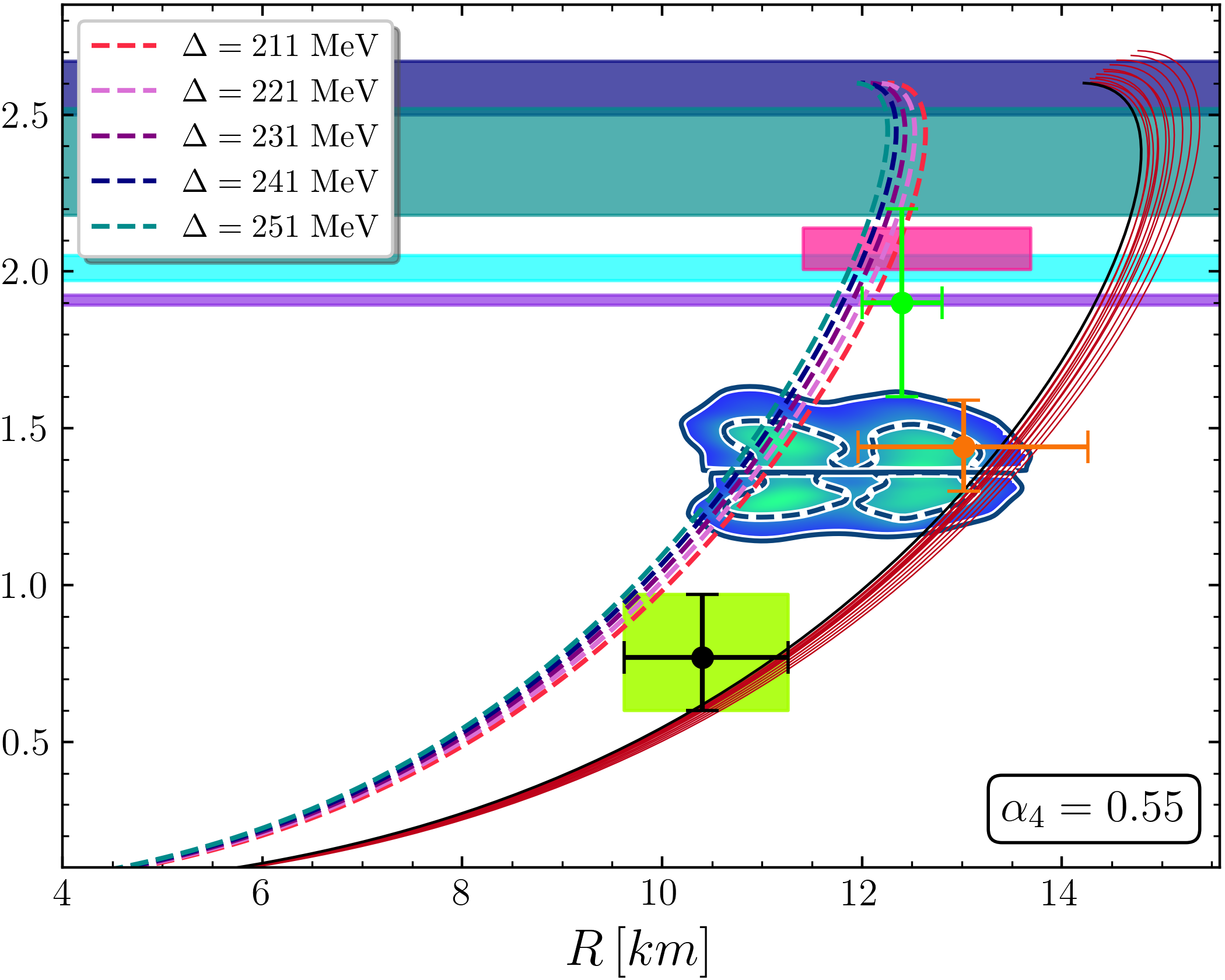}
}
\caption{Mass-Radius relations of CFL quark stars for (a) $\alpha_4=0.65$ and various $\Delta-B_{\text{eff}}$ pairs ($\Delta \in [80,180]$ MeV and $B_{\text{eff}}^{1/4} \in [130.6,161]$ MeV) that satisfy the constraints deduced in Sec.\textcolor{blue}{~\ref{sec3}}. (b) $\alpha_4=0.55$ and several $\Delta-B_{\text{eff}}$ pairs that also satisfy the aforementioned limits, with the red solid lines additionally lying within the $w \geq 0$ space ($\Delta \in [35.5,47.5]$ MeV and $B_{\text{eff}}^{1/4} \in [124.87,127.32]$ MeV), and the coloured dashed ones being consistent with the GW$170817$ and GW$190425$ tidal constraints (\{$\Delta=211$ MeV, $B_{\text{eff}}^{1/4}=169.08$ MeV\},\{$\Delta=221$ MeV, $B_{\text{eff}}^{1/4}=170.94$ MeV\},\{$\Delta=231$ MeV, $B_{\text{eff}}^{1/4}=172.66$ MeV\},\{$\Delta=241$ MeV, $B_{\text{eff}}^{1/4}=174.25$ MeV\} and \{$\Delta=251$ MeV, $B_{\text{eff}}^{1/4}=175.75$ MeV\}). The black solid line represents the softest EoS possible in the $w \geq 0$ space with $[\Delta=47.5, \, B_{\text{eff}}^{1/4}=127.32]$ MeV. From top to bottom, the coloured bands correspond to measurements of, GW$190814$'s secondary companion \textcolor{blue}{\cite{Abbott-2020a}}, PSR J$0952$-$0607$ \textcolor{blue}{\cite{Romani-2022}}, PSR J$0740$+$6620$'s marginalised posterior distribution \textcolor{blue}{\cite{Riley-2021}}, PSR J$0348$+$0432$ \textcolor{blue}{\cite{Antoniadis-2013}}, PSR J$1614$-$2230$ \textcolor{blue}{\cite{Arzoumanian-2018}}, GW$170817$ (solid and dashed lines represent the 90\% and 50\% credible level respectively) \textcolor{blue}{\cite{Abbott-2018}} and HESS J$1731$-$347$'s marginalised distribution \textcolor{blue}{\cite{Doroshenko-2022}}. The error bars with null background represent the marginalised distributions of $4U1702-429$ \textcolor{blue}{\cite{Nattila-2017}} and  PSR J$0030$+$0451$ \textcolor{blue}{\cite{Miller-2019}} correspondingly.}\label{fig7}
\end{figure*}

\begin{figure*}[t]
\subfloat[]{
\includegraphics[height=6.5cm,width=1\columnwidth]{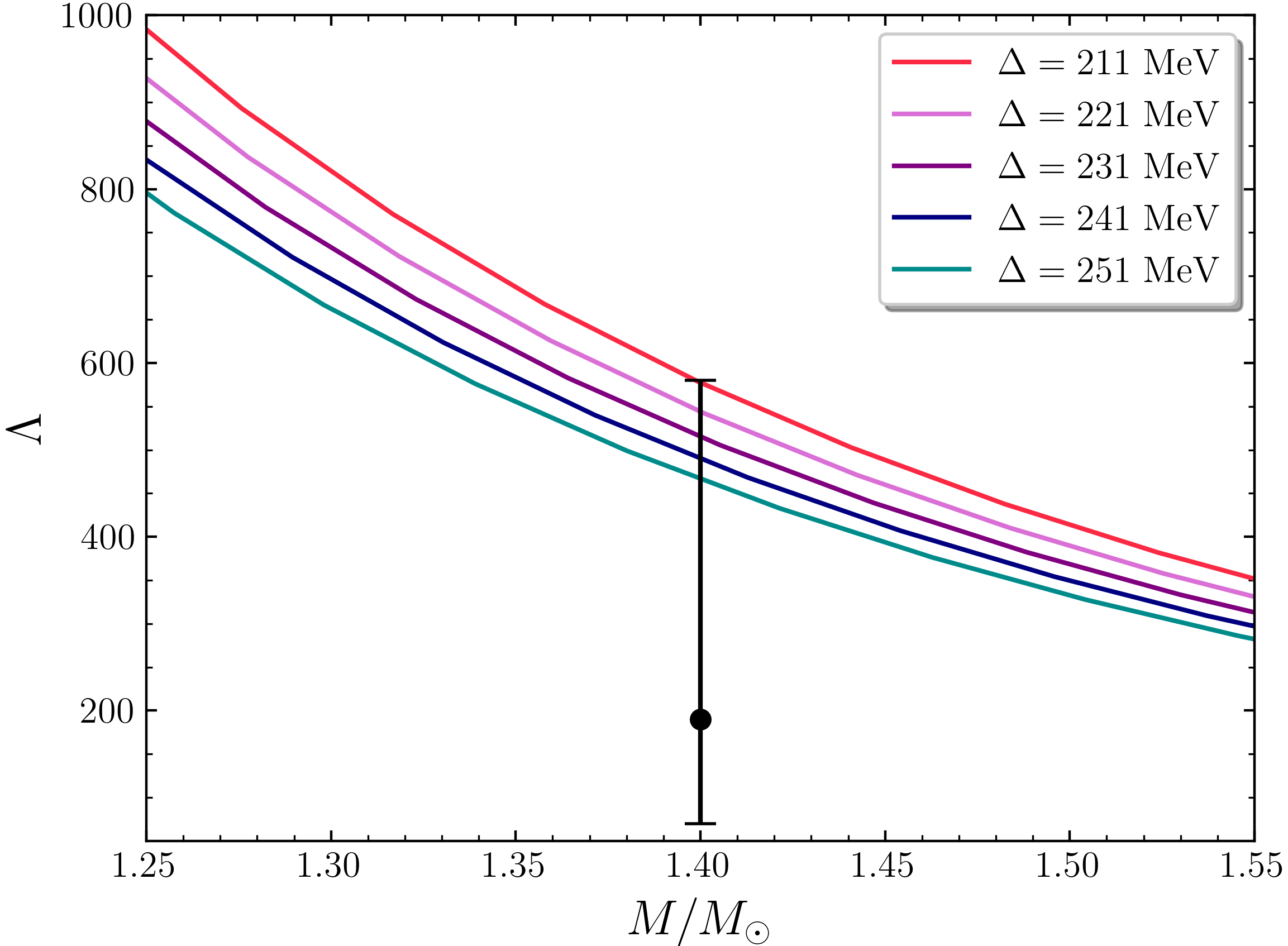}
}
\subfloat[]{
\includegraphics[height=6.5cm,width=1\columnwidth]{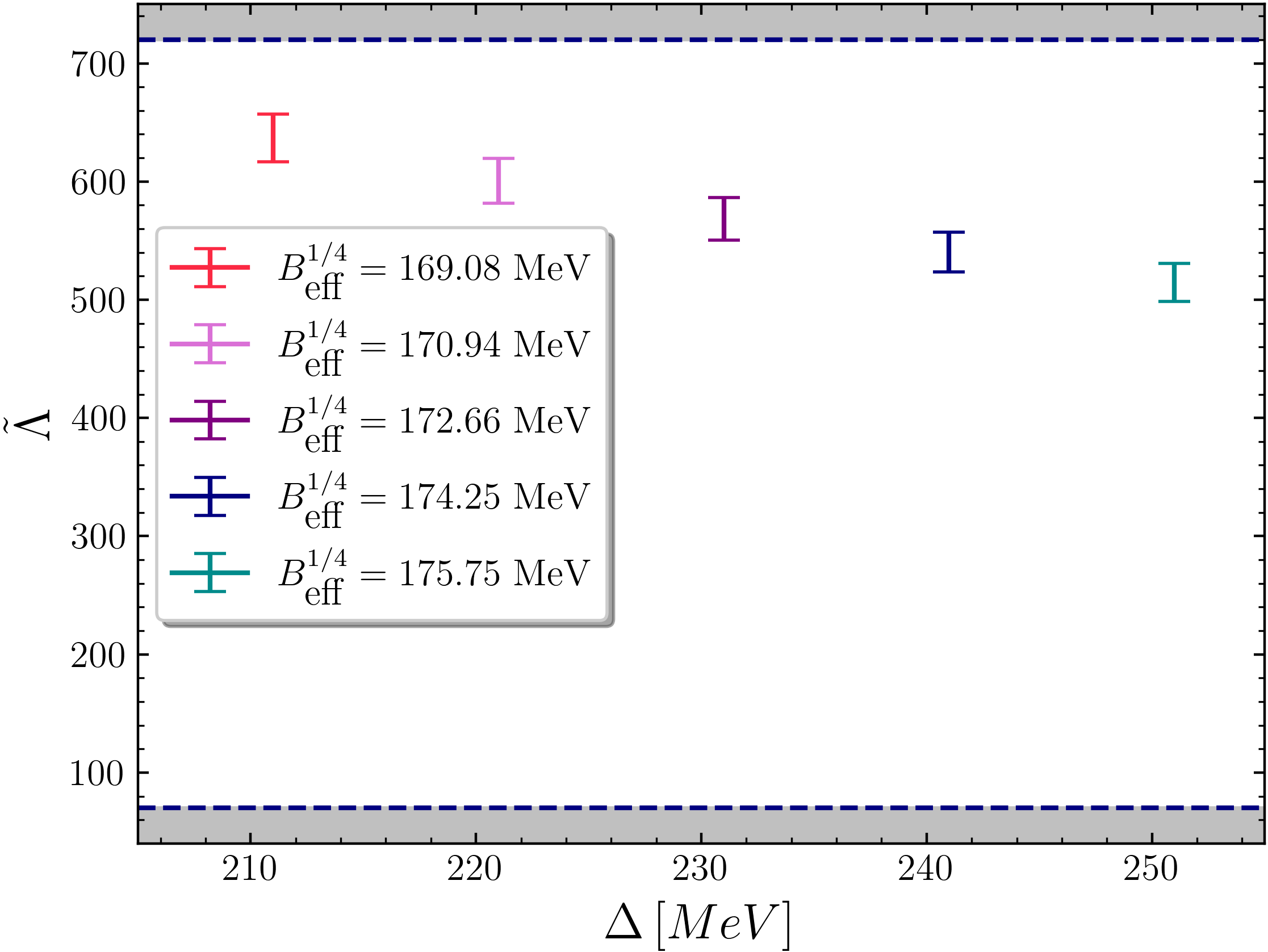}
}
\caption{The compliance of CFL quark stars with the GW$170817$ merger. (a) Dimensionless tidal deformability $\Lambda$ as a function of the normalised mass $M/M_{\odot}$, adjusted at $1.4M_{\odot}$. The error bar represents the $\Lambda_{1.4}=190^{+390}_{-120}$ constraints concluded by the LVC \textcolor{blue}{\cite{Abbott-2018}}. (b) The average tidal deformability $\Tilde{\Lambda}$ for different parametrizations. The dashed lines indicate the limits of the $\Tilde{\Lambda}=300^{+420}_{-230}$ constraint \textcolor{blue}{\cite{Abbott-2019}}. The utilised $\Delta-B_{\text{eff}}$ pairs are identical to the ones from Fig.\textcolor{blue}{~\ref{fig7}(b)}.}\label{fig8}
\end{figure*}

\begin{figure}[b]
    \includegraphics[width=245pt,height=19pc]{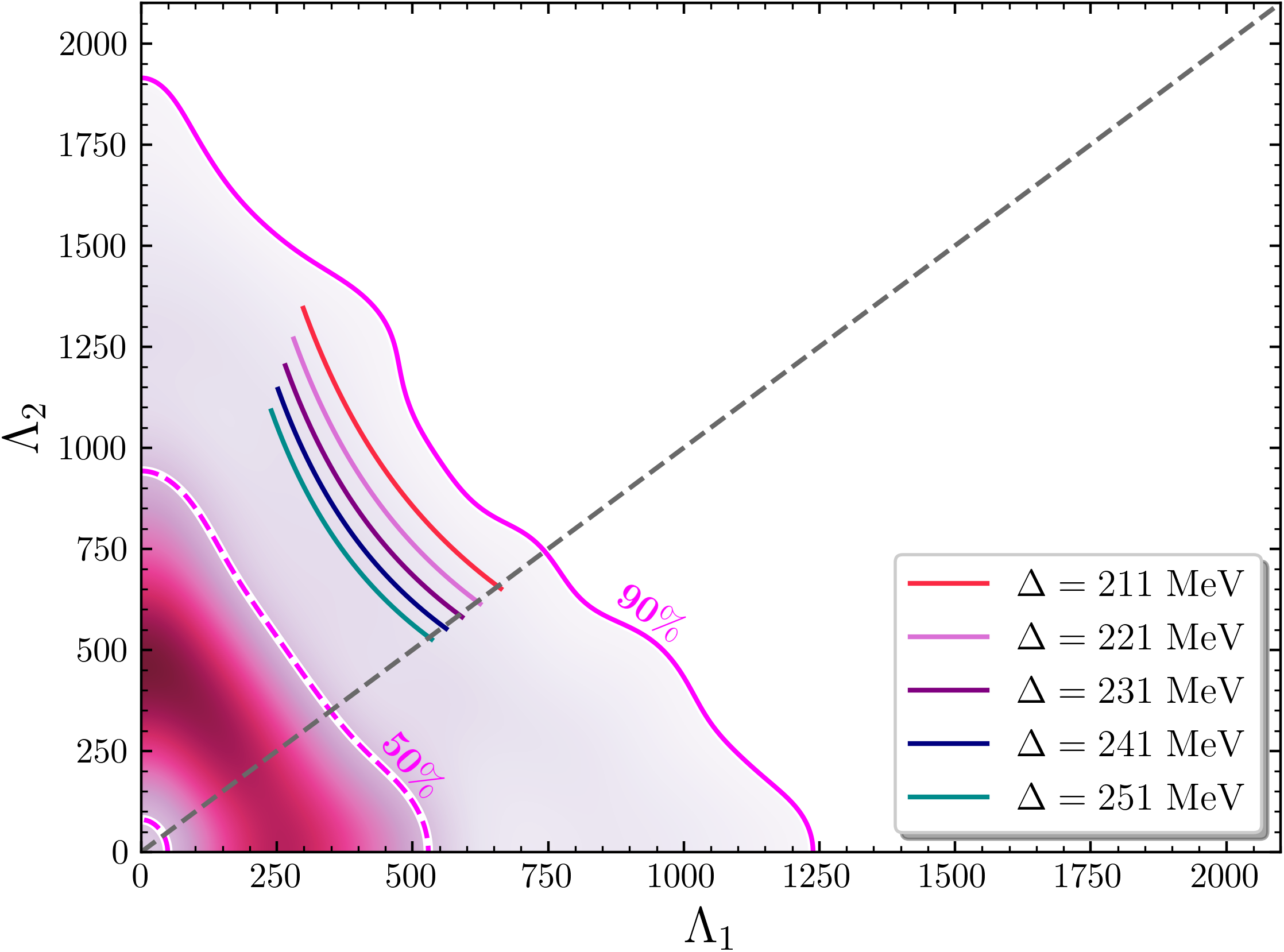}
    \caption{$\Lambda_1$-$\Lambda_2$ relation for the parameters utilised in Fig.\textcolor{blue}{~\ref{fig7}(b)} and Fig.\textcolor{blue}{~\ref{fig8}}. The $50\%$ and $90\%$ credible lines correspond to the GW$170817$ event and are obtained from \textcolor{blue}{\cite{Abbott-2018}}.}\label{fig9}
\end{figure}

\begin{figure}[b]
\includegraphics[width=245pt,height=19pc]{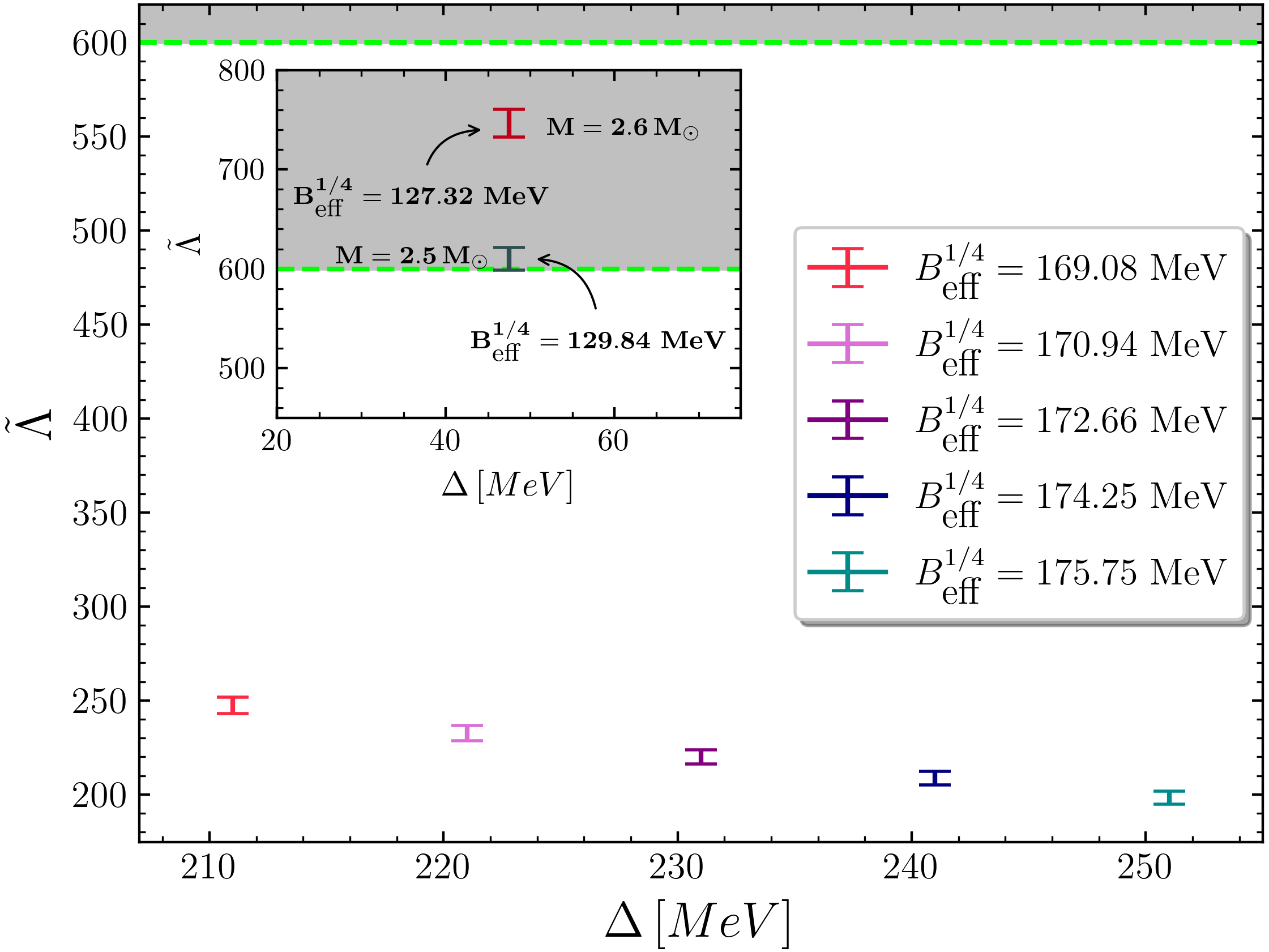}
\caption{The average tidal deformability $\Tilde{\Lambda}$ for different parametrizations. The dashed line represents the $\Tilde{\Lambda} \leq 600$ constraint of the GW$190425$ event concluded by the LVC \textcolor{blue}{\cite{Abbott-2020b}}. Parameter pairs that lie within the confined window correspond to the same pairs that were utilised in Fig.\textcolor{blue}{~\ref{fig8}}, while the two remaining pairs correspond to the softest possible EoS that utilises $w \geq 0$ parameters, for two different established $M_{\text{TOV}}=2.5$, $2.6 \, M_{\odot}$.}\label{fig10}
\end{figure}

It is concluded that in the $w<0$ space violations occur in both of the scenarios examined above, contrary to the $w \geq 0$ case. The latter range (and thus small $\Delta$) is also favoured due to high temperatures on the surface of the HESS $J1731$-$347$ remnant as explained by J.E. Horvath \textit{et al.} in \textcolor{blue}{\cite{JEHorvath-2023}} (see also the work of F. Di Clemente \textit{et al}. in \textcolor{blue}{\cite{FDiClemente-2023}}). An additional theoretical justification for the $w \geq 0$ preference also arises from the discussion in Sec. \textcolor{blue}{\ref{sec3}}. Specifically, positive $w$ values are associated with small $B_{\text{eff}}$ values, as one can see from Fig. \ref{fig3}, which favour the existence of CFL quark starts and discourage the formation of hybrid stars, as demonstrated in \textcolor{blue}{\cite{Alford-2003}}. One possible approach to thus focus in this instance is to lower the quartic coefficient $\alpha_4$ until the objects of interest can be described by gap values in the $w \geq 0$ space (we demonstrated in Sec.\textcolor{blue}{~\ref{sec3}} that as smaller values of $\alpha_4$ are examined, the minimum gap required depletes). We accordingly focus on this proposition and find that, for the constraints discussed in Sec.\textcolor{blue}{~\ref{sec3}}, the limit placed on the quartic coefficient is about $\alpha_4 \leq 0.594$, independent of the strange quark's mass value $m_s$. The sole role on the derivation of this bound lies within the $M_{\text{TOV}}\geq 2.6 M_{\odot}$ constraint as one can see from Fig.\textcolor{blue}{~\ref{fig3}}. Thus, one can establish a lower limit to gain a higher limit on $\alpha_4$ and vice versa (for $M_{\text{TOV}} \geq 2.5 \, M_{\odot}$ the limit reads $\alpha_4 \leq 0.643$). Moreover, if we only take into account the marginalised constraints of the HESS $J1731$-$347$ remnant, we find that no such limit has to be induced since we can utilise the $w \geq 0$ space for every $\alpha_4$ value discussed in Sec.\textcolor{blue}{~\ref{sec3}}. 

\section{Properties of CFL Quark Stars}\label{sec6}
It is interesting and furthermore critical to investigate the further compliance of CFL quark stars whose parameters are extracted from the confined (blue) window of Fig.\textcolor{blue}{~\ref{fig3}}, with other astrophysical objects and gravitational wave events. In the following, we take into account two different values of the quartic coefficient $\alpha_4=0.65$ and $\alpha_4=0.55$ and examine the properties of such stars. The choice of the latter value is based on the findings of \textcolor{blue}{\cite{JEHorvath-2023}} and the conclusions of Sec.\textcolor{blue}{~\ref{sec5}}, where violations on the speed of sound and the trace anomaly for $w<0$ were depicted. These findings encourage us to consider $\alpha_4$ values that allow for the utilization of the $w \geq 0$ space. The choice of the former $\alpha_4$ value is based on the results of \textcolor{blue}{\cite{Fraga-2001}}.\

In Fig.\textcolor{blue}{~\ref{fig7}(a)} we depict numerous $M-R$ graphs for the $a_4=0.65$ case with the respective parameter pairs $\Delta-B_{\text{eff}}$ being extracted from the limited area of Fig.\textcolor{blue}{~\ref{fig3}}. The maximum displayed mass is fixed at $2.8 M_{\odot}$, while the highest achievable mass for the parameters in utilisation has a magnitude of $3.89 \, M_{\odot}$. The coloured bands and error bars represent observational data corresponding to several astrophysical objects \textcolor{blue}{\cite{Doroshenko-2022,Abbott-2020a,Romani-2022,Riley-2021,Antoniadis-2013,Arzoumanian-2018,Nattila-2017,Miller-2019}}, while the 2-d posterior with credible intervals represents the constraints on the $M-R$ properties of the GW$170817$ binary system \textcolor{blue}{\cite{Abbott-2018}}. Note that every $\Delta$ value in this instance lies within the $w<0$ space as indicated in Sec.\textcolor{blue}{~\ref{sec3}}. Even though the 2-d posterior is satisfied by a number of CFL quark stars, we find that the respective EoS that describes their interior is not soft enough to satisfy the total tidal constraints of the GW$170817$ event \textcolor{blue}{\cite{Abbott-2018,Abbott-2019}}. The production of such soft EoS is not, however, unachievable; we conclude that CFL quark stars that satisfy the parameter constraints depicted in this research, can additionally satisfy the tidal constraints of the GW$170817$ event at extremely high values of the gap parameter $\Delta$, as shown in Table \ref{tab2} for three different $\alpha_4$ values. It is concluded that the minimum gap must have values that exceed $200$ MeV, with the respective minimum effective bag parameter being about $B_{\text{eff}}^{1/4}=168.94$ MeV for every $\alpha_4$. Such high values have also been deduced and/or studied in \textcolor{blue}{\cite{Miao-2021,Roupas-2021,Baym-2018}}. Once more, the reason behind such strict confinements is solely due to the $M_{\text{TOV}}\geq 2.6 M_{\odot}$ limit. Thus, one can adjust this limit to gain less or more restrictive constraints accordingly (lower mass constraints allow for bigger $B_{\text{eff}}$ values and lower but sufficiently high $\Delta$). We further note that for such high gaps, the violation of the $\langle \Theta \rangle _{\mu_B} \geq 0$ bound occurs at densities of magnitude ${\cal E}/{\cal E}_0 \approx 3.89$, depleting for increasing $\Delta$ and respective $B_{\text{eff}}$ values that satisfy the constraints mentioned in Sec.\textcolor{blue}{~\ref{sec3}}. Consequently, the satisfaction of the $\langle \Theta \rangle _{\mu_B} \geq 0$ limit poses a challenge and can induce further constraints if we wish to explain the GW$170817$ event along with masses of magnitude $M_{\text{TOV}} \geq 2.6 M_\odot$.\ 

\begin{table}[bh]
\caption{Minimum possible values of the gap $\Delta$ needed in order for CFL quark stars to additionally satisfy the total tidal constraints of the GW$170817$ event, for three different $\alpha_4$ values. These constraints are also depicted in Fig. \ref{fig8} and Fig. \ref{fig9}. The respective minimum effective bag parameter is about $B_{\text{eff}}^{1/4}=168.94$ MeV, for every $\alpha_4$.\label{tab2}}%
\tabcolsep=30pt%
\begin{ruledtabular}
\begin{tabular}{cc}
\multicolumn{1}{c}{$\alpha_4$}&
\multicolumn{1}{c}{$\Delta_{\text{min}}$ (MeV)}\\ 
\hline

$0.55$ & $210.29$ \\
$0.65$ & $218.87$ \\
$0.75$ & $226.38$  \\
\end{tabular}
\end{ruledtabular}
\end{table}

We proceed to study the compliance of CFL quark stars with the LIGO-Virgo tidal constraints \textcolor{blue}{\cite{Abbott-2018,Abbott-2019}} for $a_4=0.55$ due to the allowed utilisation of the $w \geq 0$ space. It is however evident that in the instance examined in this work, the aforementioned tidal constraints can not be described in the $w\geq 0$ space regardless the $\alpha_4$ value. In Fig.\textcolor{blue}{~\ref{fig7}(b)} we depict $M-R$ graphs of CFL quark stars described by $a_4$=0.55 (the minimum gap required to satisfy the constraints of Sec.\textcolor{blue}{~\ref{sec3}} in this case is about $\Delta_{\text{min}}=35.01$ MeV) in the $w \geq 0$ space (the highest presented and achievable mass has a value of $2.7 \, M_{\odot}$) and for large values of the gap $\Delta$ that can satisfy the tidal constraints of GW$170817$. For the latter case, we chose the biggest possible $B_{\text{eff}}$ value for the corresponding gap $\Delta$ that allows for $M=2.6 M_{\odot}$ in order to achieve the softest EoS and thus the best correspondence with the GW$170817$ constraints. Note that in the case of $w>0$ (red solid lines) the compliance in respect to other astrophysical objects is not as robust as in the case of Fig.\textcolor{blue}{~\ref{fig7}(a)}. That is due to the narrowed $B_{\text{eff}}$ space induced by the $M_{\text{TOV}} \geq 2.6 M_{\odot}$ limit, forbidding big effective bag values and thus the further softness of the EoS. One can deal with such an inconvenience by applying bigger $\Delta$ values and extending the $B_{\text{eff}}$ window but entering the $w<0$ space. Another possible approach without the need for much bigger $\Delta$ is to establish higher $\alpha_4$ values. In this case, the EoS becomes stiffer for a given $\Delta -B_{\text{eff}}$ pair as predicted from Fig.\textcolor{blue}{~\ref{fig5}}, extending the $B_{\text{eff}}$ window and thus allowing for higher compactness. Furthermore, the observational constraints become wider as the quartic coefficient increases, supporting the aforementioned extension. However, the range between $\alpha_4=0.55$ and $\alpha_4=0.594$ is too frail for significant changes to be induced and the gap window in the $w \geq 0$ space becomes significantly more limited as the quartic coefficient rises. Additionally, we find that the softest possible EoS in the $w \geq 0$ space for an established $M_{\text{TOV}}$ occurs for $w=0$, independently of the $\alpha_4$ value. The respective bag parameter is defined by the second stability constraint (examined in Sec. \ref{sec3}) for the maximum possible $\alpha_4$ that allows for the utilisation of the $w \geq 0$ space. E.g., for $M_{\text{TOV}}=2.6 M_{\odot}$, $B_{\text{eff}}$ is defined by the second stability constraint for $\alpha_4=0.594$. The $M-R$ graph of the softest EoS is represented by the black solid line in Fig.\textcolor{blue}{~\ref{fig7}(b)}. Consequently, one can achieve a softer EoS and a more robust compliance with the depicted objects only by inducing a depletion on the $M_{\text{TOV}}$ limit.\

It is evident that the presence of an EoS softness bound in the $w \geq 0$ space implies a minimum possible value on the tidal deformability $\Lambda$ at a fixed mass $M$. For the EoS corresponding to the solid black line in Fig.\textcolor{blue}{~\ref{7}(b)}, the tidal deformability of a $2.6 M_{\odot}$ star is found to be $\Lambda_{2.6}=22.11$. The stiffest EoS presented in the same figure for $w \geq 0$, yields a value of $\Lambda_{2.6}=49.81$. However, allowing for parameter pairs from the $w <0$ space can result in magnitudes as low as $\Lambda_{2.6}=5.96$. This value corresponds to the pair that produces the softest EoS presented in this work, namely \{$\Delta=251$ MeV, $B_{\text{eff}}^{1/4}=175.75$ MeV\}. Since there were no evidence of measurable tidal effects in the GW$190814$ signal \textcolor{blue}{\cite{Abbott-2020a}}, one can compare our results with the conclusions drawn in \textcolor{blue}{\cite{Miao-2021}}, which read $4.6< \Lambda_{2.6} <26.3$.\

In Fig.\textcolor{blue}{~\ref{fig8}} and Fig.\textcolor{blue}{~\ref{fig9}} we present the compliance of CFL quark stars with the GW$170817$ event for $\alpha_4=0.55$ (for the depiction of the $M-R$ and tidal credible levels of the aforementioned event we utilised the publicly-open python functions provided by B. Lackey \textcolor{blue}{\cite{LackeyGithub-2018}}). It is worth noting that a reverse behaviour is about to take place when higher $\Delta$ values are utilised. Although extremely high gaps were needed to achieve a soft enough EoS to describe the aforementioned merger, it is evident from the figure that the range between tidal deformabilities becomes smaller and smaller as the gap rises. This indicates a maximum possible value for $\Lambda_{1.4}$ and a reverse effect for much higher $\Delta$ values. This specific behaviour is due to the constraints of the compact object in HESS J$1731$-$347$. Particularly, the softness of the EoS at such high $\Delta$ originates from the respective high $B_{\text{eff}}$ values which also bring the $M-R$ graphs close to the upper limit of the aforementioned object, as shown in  Fig.\textcolor{blue}{~\ref{fig7}(b)}. Consequently, in order to prevent the violation of that limit, the $B_{\text{eff}}$ space becomes narrower for much greater $\Delta$ as the latter quantity rises, making the respective EoS stiffer than the former one. This behaviour is predicted from Fig.\textcolor{blue}{~\ref{fig3}} where the $R_{0.9\text{min}}$ constraint steeply approaches the $M_{\text{TOV}}$ limit. For $\alpha_4=0.55$, the lines meet at about $\Delta \approx 253$ MeV. Of course, one can deal with wider posterior distributions (e.g. $1 \sigma$ or $2 \sigma$ confidence levels) to avoid such circumstances.\

Lastly, in Fig.\textcolor{blue}{~\ref{fig10}} we show the behaviour of CFL quark stars in respect to the GW$190425$ event \textcolor{blue}{\cite{Abbott-2020b}}. The two parameter pairs that lie outside the required $\Tilde{\Lambda}$ window correspond to the softest possible EoS for $w \geq 0$ and $M_{\text{TOV}}=2.5 ,2.6 \, M_{\odot}$, while the rest included parameter pairs are those utilised in Fig.\textcolor{blue}{~\ref{fig8}}. We thus conclude that the equations of state corresponding to the $w \geq 0$ do not comply well with the GW$190425$ event either. We also inform the reader that the absence of an equal mass bound can induce some modest errors on our results.
\section{Conclusions}\label{sec7}
 In this study, we aimed to explain the central compact object within the HESS J$1731$-$347$ remnant as a CFL quark star by using its marginalised posterior distribution and imposing it as a constraint on the relevant parameter space. We also required the quark stars to have sufficiently high masses ($\geq 2.6 \, M_{\odot}$) to account for GW$190814$'s secondary companion.
By examining their behaviour in
respect to the speed of sound and the trace anomaly, we concluded that violations occur in both cases for $w < 0$, whereas for $w \geq 0$ the contradictory takes effect. This indicates that their interior strongly depends on the sign of the parameter  $w$ ($a_2$), which was previously noted by Miao $\textit{et al}.$ for the case of the speed of sound \textcolor{blue}{\cite{Miao-2021}} (for further discussion see the work of Alford \textit{et al.} \textcolor{blue}{\cite{Alford-2005}}). Despite the need for extremely stiff equations of state to reach masses of $2.6 \, M_{\odot}$, we showed that $w<0$ values are not necessary to describe the compact objects of interest, as long as the quartic coefficient satisfies $\alpha_4 \leq 0.594$. Indicating that we can attain high masses without violating the conformal limit or the positive trace anomaly bound. However, these quartic coefficient values deviate from the expected value of $\approx 0.65$ \textcolor{blue}{\cite{Fraga-2001,Alford-2005}} and will diverge further if higher mass limits are established.  Conversely, we can achieve values closer to the expected one if a lower mass limit is taken into account (we found that for $M_{\text{TOV}} \geq 2.5 M_{\odot}$ the limit becomes $\alpha_4 \leq 0.643$). In the instance of neglecting an $M_{\text{TOV}}$ bound, we can explain the compact object in HESS J$1731$-$347$ for any value of $a_4$ in the examined range ($[0.45,1]$) without exceeding the $w>0$ area. The utilisation of the $w \geq 0$ space and thus small $\Delta$ is not only favoured due to the evasion of the aforementioned violations, but additionally due to the concluded surface temperatures of the latter object, as presented in \textcolor{blue}{\cite{JEHorvath-2023}}, and the restriction of the $B_{\text{eff}}$ to small values which favours the existence of CFL quark stars \textcolor{blue}{\cite{Alford-2003}}. The insertion to the $w<0$ is, however, inevitable if we were also to account for the GW$170817$ tidal constraints, which require extremely high gaps ($\Delta>200$ MeV) and thus cannot be described in the $w \geq 0$ space. Such magnitudes though can be decreased if we establish a lower mass limit which results in softer EoS. This additionally favours the compliance of CFL quark stars described in the $w \geq 0$ space with other various astrophysical objects, as elaborated in Sec.\textcolor{blue}{~\ref{sec6}}. Our findings also reveal a softness bound in the $w \geq 0$ space, implying a lower limit on the tidal deformability $\Lambda$ at a fixed mass $M$. Specifically, the lowest achievable value in the $w \geq 0$ range for a $2.6 \, M_{\odot}$ star is found to be $\Lambda_{2.6}=22.11$. However, by considering parameter pairs from the $w<0$ space, we can obtain significantly lower magnitudes, with the softest equation of state (EoS) presented in this research yielding $\Lambda_{2.6}=5.96$. These values can be compared with the confinements concluded in \textcolor{blue}{\cite{Miao-2021}} which regard the GW$190814$ event and read $4.6< \Lambda_{2.6} <26.3$. Future observations and studies will shed light on the objects mentioned and examined in this work, further determining the course of our findings. Lastly, it is worth noting that in the current study we considered non-rotating quark stars. However, exploring the scenario of rapid rotation and comparing the results with those of static configurations and existing observational events holds great significance. A study investigating this aspect is currently underway.\\


\section*{Acknowledgments}
The authors thank  Prof. L. Rezzolla for the useful discussion and comments along with A. Kanakis-Pegios for his valuable contribution regarding the GW$170817$ event.


\begin{thebibliography}{85}

\bibitem{Witten-1984}
E. Witten, \href{https://journals.aps.org/prd/abstract/10.1103/PhysRevD.30.272}{Phys. Rev. D 30, 272 (1984)}.

\bibitem{Bodmer-1971}
A. R. Bodmer, \href{https://journals.aps.org/prd/abstract/10.1103/PhysRevD.4.1601}{Phys. Rev. D 4, 1601 (1971)}.

\bibitem{Ivanenko-1965}
D. D. Ivanenko and D. F. Kurdgelaidze, \href{https://link.springer.com/article/10.1007/BF01042830}{ Astrophysics 1, 251 (1965)}.

\bibitem{Itoh-1970}
 N. Itoh, \href{https://academic.oup.com/ptp/article/44/1/291/1896619}{Prog. Theor. Phys. 44, 291 (1970)}.

\bibitem{Collins-1975}
J.C. Collins, and M.J. Perry ,\href{https://journals.aps.org/prl/abstract/10.1103/PhysRevLett.34.1353}{Phys. Rev. Lett. 34, 1353 (1975)}.


\bibitem{Weber-2005} F.Weber,\href{https://www.sciencedirect.com/science/article/pii/S0146641004001061?via%3Dihub}{Prog. Part. Nucl. Phys. 54, 193 (2005)}.

\bibitem{Bielich-2020} J. Schaffner-Bielich, \textit{Compact Star Physics} (Cambridge University Press, Cambridge, England,2020).

\bibitem{Glendenning-2000}
 N. K. Glendenning, Compact Stars: Nuclear Physics, Particle
Physics, and General Relativity, 2nd ed. (Springer, New York,
2000)

\bibitem{Alock-1986}
 C. Alcock, E. Farhi, and A. Olinto, \href{https://ui.adsabs.harvard.edu/abs/1986ApJ...310..261A/abstract}{Astrophys. J. 310, 261
(1986)}.

\bibitem{Hanesel-1986}
 P. Haensel, J. L. Zdunik, and R. Schaeffer, \href{https://adsabs.harvard.edu/full/1986A%26A...160..121H}  {Astron. Astrophys. 160, 121 (1986).}



\bibitem{Alford-2001a}
 M. G. Alford, \href{https://www.annualreviews.org/doi/10.1146/annurev.nucl.51.101701.132449#_i15}{Annu. Rev. Nucl. Part. Sci. 51, 131 (2001)}.

 \bibitem{Alford-2008}
 M. G. Alford, A. Schmitt, K. Rajagopal, and T. Schäfer, \href{https://journals.aps.org/rmp/abstract/10.1103/RevModPhys.80.1455}{Rev. Mod. Phys. 80, 1455 (2008)}.

 \bibitem{Rajagopal-2000}
 K. Rajagopal and F. Wilczek, The condensed matter
physics of QCD, in At the Frontier of Particle
Physics. Handbook of QCD, edited by M. Shifman and
B. Ioffe, Vol. 1–3 (World Scientific, Singapore, 2000),
pp. 2061–2151.

\bibitem{Alford-1999}
 M. G. Alford, K. Rajagopal, and F. Wilczek, \href{https://www.sciencedirect.com/science/article/abs/pii/S0550321398006683?via%3Dihub}{Nucl. Phys.
B537, 443 (1999)}.

\bibitem{Alford-2002}
M. Alford and K. Rajagopal, \href{https://iopscience.iop.org/article/10.1088/1126-6708/2002/06/031}{J. High Energy Phys. 06
(2002) 031}.

\bibitem{Doroshenko-2022}
 V. Doroshenko, V. Suleimanov, G. Phlhofer, and Andrea
Santangelo, \href{https://www.nature.com/articles/s41550-022-01800-1}{Nat. Astron. 6, 1444 (2022)}.

\bibitem{Suwa-2018}
Y. Suwa, T. Yoshida, M. Shibata, H. Umeda, and K.
Takahashi, \href{https://academic.oup.com/mnras/article/481/3/3305/5094598}{Mon. Not. R. Astron. Soc. 481, 3305 (2018)}.

\bibitem{Tsaloukidis-2023}
L. Tsaloukidis, P.S. Koliogiannis, A. Kanakis-Pegios, and Ch.C. Moustakidis, \href{https://journals.aps.org/prd/abstract/10.1103/PhysRevD.107.023012}{Phys. Rev. D 107, 023012 (2023)}.

\bibitem{Brodie-2023}
L. Brodie, and A. Haber, \href{https://arxiv.org/abs/2302.02989}{arXiv:2302.02989 (2023)}.

\bibitem{JEHorvath-2023}
J.E. Horvath, L.S. Rocha, L.M. de Sá, P.H.R.S. Moraes, L. G. Barão, M.G.B. de Avellar, A. Bernardo, and R.R.A. Bachega, \href{https://arxiv.org/abs/2303.10264}{arXiv:2303.10264 (2023)}.

\bibitem{FDiClemente-2023}
F. Di Clemente, A. Drago, and G. Pagliara, \href{https://arxiv.org/abs/2211.07485}{arXiv:2211.07485 (2023)}.

\bibitem{Abbott-2020a}
R. Abbott et al. (LIGO Scientific, Virgo Collaborations), \href{https://iopscience.iop.org/article/10.3847/2041-8213/ab960f}{Astrophys. J. Lett. 896, L44 (2020)}.

\bibitem{Chodos-1974}
A. Chodos, R. L. Jaffe, K. Johnson, C. B. Thorn, and V. F. Weisskopf \href{https://journals.aps.org/prd/abstract/10.1103/PhysRevD.9.3471}{Phys. Rev. D 9, 3471 (1974)}.

\bibitem{Fujimoto-2022}
Y. Fujimoto, K. Fukushima, L. D. McLerran, and M. Praszałowicz \href{https://journals.aps.org/prl/abstract/10.1103/PhysRevLett.129.252702}{Phys. Rev. Lett. 129, 252702 (2022)}.

\bibitem{GLugones-2003}
G. Lugones and J. E. Horvath, \href{https://www.aanda.org/articles/aa/abs/2003/19/aa3362/aa3362.html}{Astron. Astrophys. 403, 173
(2003)}.

\bibitem{JEHorvath-2004}
 J. E. Horvath and G. Lugones,\href{https://www.aanda.org/articles/aa/abs/2004/28/aagb161/aagb161.html}{Astron. Astrophys. 422, L1
(2004)}.

\bibitem{AKurkela-2010}
A. Kurkela, P. Romatschke, and A. Vuorinen, \href{https://journals.aps.org/prd/abstract/10.1103/PhysRevD.81.105021}{Phys. Rev. D 81, 105021 (2010)}.

\bibitem{CVFlores-2017}
C. V. Flores and G. Lugones, \href{https://journals.aps.org/prc/abstract/10.1103/PhysRevC.95.025808}{Phys. Rev. C 95, 025808
(2017)}.

\bibitem{Romani-2022}
R. W. Romani, D. Kandel, A. V. Filippenko, T. G. Brink, and
W. Zheng, \href{https://iopscience.iop.org/article/10.3847/2041-8213/ac8007}{Astrophys. J. Lett. 934, L17 (2022)}.

\bibitem{Riley-2021}
T. E. Riley \textit{et al}.,\href{https://iopscience.iop.org/article/10.3847/2041-8213/ac0a81}{Astrophys. J. Lett. 918, L27 (2021)}.

\bibitem{Antoniadis-2013}
J. Antoniadis, P. Freire, N. Wex \textit{et al}., \href{https://www.science.org/doi/10.1126/science.1233232}{Science 340, 448
(2013)}.

\bibitem{Arzoumanian-2018}
Z. Arzoumanian, A. Brazier, S. Burke-Spolaor \textit{et al}., \href{https://iopscience.iop.org/article/10.3847/1538-4365/aab5b0}{Astrophys. J. Suppl. Ser. 235, 37 (2018)}.

\bibitem{Nattila-2017}
 J. Nättilä, M. C. Miller, A. W. Steiner, J. J. E. Kajava, V. F.
Suleimanov, and J. Poutanen, \href{https://www.aanda.org/articles/aa/full_html/2017/12/aa31082-17/aa31082-17.html}{Astron. Astrophys. 608, A31 (2017)}.


\bibitem{Miller-2019}
M. C. Miller \textit{et al}., \href{https://iopscience.iop.org/article/10.3847/2041-8213/ab50c5}{Astrophys. J. Lett. 887, L24 (2019)}.

\bibitem{Abbott-2017}
 B. P. Abbott \textit{et al}., (LIGO Scientific and Virgo Collaborations), \href{https://journals.aps.org/prl/abstract/10.1103/PhysRevLett.119.161101}{Phys. Rev. Lett. 119, 161101 (2017)}.

\bibitem{Abbott-2020b}
B. P. Abbott \textit{et al}., (LIGO Scientific and Virgo Collaborations), \href{https://iopscience.iop.org/article/10.3847/2041-8213/ab75f5}{Astrophys. J. Lett. 892, L3 (2020)}.


\bibitem{Alford-2003}
M. Alford and S. Reddy, \href{https://journals.aps.org/prd/abstract/10.1103/PhysRevD.67.074024}{Phys. Rev. D 67, 074024 (2003)}.

\bibitem{Alford-2005}
M. Alford, M. Braby, M. W. Paris, and S. Reddy, \href{https://iopscience.iop.org/article/10.1086/430902}{Astrophys.
J. 629, 969 (2005)}.

\bibitem{Rajagopal-2001}
K. Rajagopal and F. Wilczek, \href{https://journals.aps.org/prl/abstract/10.1103/PhysRevLett.86.3492}{Phys. Rev. Lett. 86, 3492 (2001)}.

\bibitem{Workman-2022}
R. L. Workman \textit{et al}. (Particle Data Group), \href{https://academic.oup.com/ptep/article/2022/8/083C01/6651666}{Prog.
Theor. Exp. Phys. 2022, 083C01}. 

\bibitem{Fraga-2001}
E. S. Fraga, R. D. Pisarski, and J. Schaffner-Bielich, \href{https://journals.aps.org/prd/abstract/10.1103/PhysRevD.63.121702}{Phys. Rev. D 63, 121702 (2001)}.


\bibitem{Weissenborn-2011}
 S. Weissenborn, I. Sagert, G. Pagliara, M. Hempel, and J.
Schaffner-Bielich, \href{https://iopscience.iop.org/article/10.1088/2041-8205/740/1/L14}{Astrophys. J. Lett. 740, L14 (2011)}.

\bibitem{Alford-2001b}
M. Alford, K. Rajagopal, S. Reddy, and F. Wilczek, \href{https://journals.aps.org/prd/abstract/10.1103/PhysRevD.64.074017}{Phys. Rev. D 64, 074017 (2001)}.

\bibitem{Lugones-2002}
G. Lugones and J. E. Horvath, \href{https://journals.aps.org/prd/abstract/10.1103/PhysRevD.66.074017}{Phys. Rev. D 66, 074017 (2002)}.

\bibitem{Zhang-2021}
C. Zhang and R. B. Mann, \href{https://journals.aps.org/prd/abstract/10.1103/PhysRevD.103.063018}{Phys. Rev. D 103, 063018 (2021)}.

\bibitem{Alford-2004}
M. Alford, C. Kouvaris, and K. Rajagopal, \href{https://journals.aps.org/prl/abstract/10.1103/PhysRevLett.92.222001}{Phys. Rev. Lett. 92, 222001 (2004)}.


\bibitem{Farhi-1984}
E. Farhi and R. L. Jaffe, \href{https://journals.aps.org/prd/abstract/10.1103/PhysRevD.30.2379}{Phys. Rev. D 30, 2379 (1984)}.

\bibitem{Zhou-2018}
E.-P. Zhou, X. Zhou, and A. Li, \href{https://journals.aps.org/prd/abstract/10.1103/PhysRevD.97.083015}{Phys. Rev. D 97, 083015 (2018)}.

\bibitem{Miao-2021}
 Z. Miao, J.-L. Jiang, A. Li, and L.-W. Chen, \href{https://iopscience.iop.org/article/10.3847/2041-8213/ac194d}{Astrophys. J.
917, L22 (2021)}.

\bibitem{Buballa-1999}
M. Buballa and M. Oertel, \href{https://www.sciencedirect.com/science/article/pii/S037026939900533X?via%3Dihub}{Phys. Lett. B 457, 261
(1999)}.

\bibitem{Holdom-2018}
 B. Holdom, J. Ren, and C. Zhang, \href{https://journals.aps.org/prl/abstract/10.1103/PhysRevLett.120.222001}{Phys. Rev. Lett. 120,
222001 (2018)}.

\bibitem{Alford-1999b}
M. Alford, J Berges, and K. Rajagopal, \href{https://www.sciencedirect.com/science/article/pii/S0550321399004101?via%3Dihub}{Nucl. Phys. B558, 219 (1999)}.

\bibitem{Schafer-2000}
T. Schäfer, \href{https://www.sciencedirect.com/science/article/pii/S0550321300000638?via%3Dihub}{Nucl. Phys. B575, 269 (2000)}.

\bibitem{QWang-2020}
 Q. Wang, T. Zhao, and H. Zong, \href{https://www.worldscientific.com/doi/abs/10.1142/S0217732320503216}{Mod. Phys. Lett. A 35,
2050321 (2020)}.

\bibitem{Bombaci-2021}
I. Bombaci, A. Drago, D. Logoteta, G. Pagliara, and I. Vidaña, \href{https://journals.aps.org/prl/abstract/10.1103/PhysRevLett.126.162702}{Phys. Rev. Lett. 126, 162702 (2021)}.

\bibitem{Postnikov-2010}
S. Postnikov, M. Prakash, and J. M. Lattimer, \href{https://journals.aps.org/prd/abstract/10.1103/PhysRevD.82.024016}{Phys. Rev. D 82, 024016 (2010)}.

\bibitem{Flanagan-2008}
E. E. Flanagan and T. Hinderer, \href{https://journals.aps.org/prd/abstract/10.1103/PhysRevD.77.021502}{Phys. Rev. D 77, 021502(R)
(2008)}.

\bibitem{Hinderer-2008}
T. Hinderer, \href{https://iopscience.iop.org/article/10.1086/533487}{Astrophys. J. 677, 1216 (2008)}.

\bibitem{Binnington-2009}
T. Binnington and E. Poisson, \href{https://journals.aps.org/prd/abstract/10.1103/PhysRevD.80.084018}{Phys. Rev. D 80, 084018 (2009)}.

\bibitem{Damour-1992}
T. Damour, M. Soffel, and C. Xu, \href{https://journals.aps.org/prd/abstract/10.1103/PhysRevD.45.1017}{Phys. Rev. D 45, 1017 (1992)}.

\bibitem{Damour-2009}
T. Damour and A. Nagar, \href{https://journals.aps.org/prd/abstract/10.1103/PhysRevD.80.084035}{Phys. Rev. D 80, 084035 (2009)}.


\bibitem{Hinderer-2010}
T. Hinderer, B. D. Lackey, R. N. Lang, and J. S. Read, \href{https://journals.aps.org/prd/abstract/10.1103/PhysRevD.81.123016}{Phys. Rev. D 81, 123016 (2010)}.

\bibitem{Li-2021}
B.-L. Li, Y. Yan, and J.-L. Ping,
\href{https://journals.aps.org/prd/abstract/10.1103/PhysRevD.104.043002}{Phys. Rev. D 104, 043002 (2021)}.

\bibitem{Lourenco-2021}
O. Lourenço, C. H. Lenzi, M. Dutra, E. J. Ferrer, V. de la
Incera, L. Paulucci, and J. E. Horvath, \href{https://journals.aps.org/prd/abstract/10.1103/PhysRevD.103.103010}{Phys. Rev. D 103, 103010 (2021)}.

\bibitem{Bedaque-2015}
P. Bedaque and A. W. Steiner, \href{https://journals.aps.org/prl/abstract/10.1103/PhysRevLett.114.031103}{Phys. Rev. Lett. 114, 031103 (2015)}.

\bibitem{Tews-2018}
I. Tews, J. Carlson, S. Gandolfi, and S. Reddy, \href{https://iopscience.iop.org/article/10.3847/1538-4357/aac267}{Astrophys. J.
860, 149 (2018)}.

\bibitem{Traversi-2022}
S. Traversi, P. Char, G. Pagliara, and A. Drago, \href{https://www.aanda.org/articles/aa/full_html/2022/04/aa41544-21/aa41544-21.html}{Astron. Astrophys. 660, A62 (2022)}.

\bibitem{Moustakidis-2017}
C. C. Moustakidis, T. Gaitanos, C. Margaritis, and G. A. Lalazissis, \href{https://journals.aps.org/prc/abstract/10.1103/PhysRevC.95.045801}{Phys. Rev. C 95, 045801 (2017)}.

\bibitem{Roupas-2021}
Z. Roupas, G. Panotopoulos, and I. Lopes, \href{https://journals.aps.org/prd/abstract/10.1103/PhysRevD.103.083015}{Phys. Rev. D 103, 083015 (2021)}.

\bibitem{Ecker-2022} C. Ecker and L. Rezzolla, \href{https://doi.org/10.3847/2041-8213/ac8674
}{Astrophys. J.
939, L35 (2022)}.

\bibitem{Altiparmak-2022}S. Altiparmak,  C. Ecker,  and L. Rezzolla, \href{
https://doi.org/10.3847/2041-8213/ac9b2a} {Astrophys. J.
939, L34 (2022)}.

\bibitem{Hippert-2021}M. Hippert, E.S. Fraga, and J. Noronha, \href{
https://doi.org/10.1103/PhysRevD.104.034011
}{Phys. Rev. D 104, 034011 (2021)}.

\bibitem{Blaschke-2022} O. Ivanytskyi and D.B. Blaschke, \href{
https://doi.org/10.3390/particles5040038
}{Particles 5, 514-524 (2022)}. 

\bibitem{Annala-2023} E. Annala, T. Gorda, J. Hirvonen, O. Komoltsev, A. Kurkela, J. Nättilä, and A. Vuorinen, \href{https://arxiv.org/abs/2303.11356}{	arXiv:2303.11356 (2023)}. 


\bibitem{Comment-1}
A straightforward way to verify this is with a simple differentiation on the EoS deduced in \textcolor{blue}{\cite{Zhang-2021}} for $\lambda<0$.

\bibitem{Marczenko-2023} M. Marczenko, L. McLerran, K. Redlich, and C. Sakaki,  \href{https://journals.aps.org/prc/abstract/10.1103/PhysRevC.107.025802} {Phys. Rev. C 107, 025802 (2023)}.

\bibitem{Abbott-2018}
B.P. Abbott \textit{et al}. (The LIGO Scientific Collaboration and the Virgo Collaboration), \href{https://journals.aps.org/prl/abstract/10.1103/PhysRevLett.121.161101}{Phys. Rev. Lett. 121, 161101 (2018)}.

\bibitem{Abbott-2019}
B.P. Abbott \textit{et al}. (The LIGO Scientific Collaboration and the Virgo Collaboration), \href{https://journals.aps.org/prx/abstract/10.1103/PhysRevX.9.011001}{Phys. Rev. X 9, 011001 (2019)}.

\bibitem{Baym-2018}
 G. Baym, T. Hatsuda, T. Kojo, P. D. Powell, Y. Song, and T.
Takatsuka, \href{https://iopscience.iop.org/article/10.1088/1361-6633/aaae14}{Rep. Prog. Phys. 81, 056902 (2018)}.

\bibitem{LackeyGithub-2018}
B. Lackey, GitHub repository (2018), \href{https://github.com/benjaminlackey/eosinference}{https://github.com/
benjaminlackey/eosinference.git}.



\end{thebibliography}
\end{document}